\documentclass[reprint,twocolumn,amsmath,amssymb,aps,
]{revtex4-1}

\usepackage{epsfig,bm,epsf,graphics}
\usepackage{graphicx}
\usepackage{dcolumn}
\usepackage{bm}

\begin{document}
\preprint{APS/123-QED}

\title{Magnetic properties of perovskites La$_{0.7}$Sr$_{0.3}$Mn$_{0.7}^{3+}$Mn$_{0.3-x}^{4+}$Ti$_{x}$O$_{3}$:
Monte Carlo simulation versus experiments}

\author{Samia Yahyaoui\footnote{yahyaoui.samia@yahoo.fr}}
\affiliation{%
Unit\'e de Recherche de Physique Quantique, D\'epartement de Physique,
Facult\'e des sciences de Monastir, BP 22, 5019 Monastir, Tunisia.\\
 }%


\author{Sami Kallel\footnote{sami\_kallel@yahoo.fr}}
\affiliation{%
Laboratoire de Physico-chimie des Mat\'eriaux, D\'epartement de Physique,
Facult\'e des sciences de Monastir, BP 22, 5019 Monastir, Tunisia.\\
 }%

\author{H. T. Diep\footnote{diep@u-cergy.fr}}
\affiliation{%
Laboratoire de Physique Th\'eorique et Mod\'elisation,
Universit\'e de Cergy-Pontoise, CNRS, UMR 8089\\
2, Avenue Adolphe Chauvin, 95302 Cergy-Pontoise Cedex, France.\\
}%


\date{\today}

\begin{abstract}

This work presents a Monte Carlo study of the phase transition in  the perovskites  La$_{0.7}$Sr$_{0.3}$Mn$_{0.7}^{3+}$Mn$_{0.3-x}^{4+}$Ti$_{x}$O$_{3}$ ($x$= 0.1, 0.2, and 0.25).  We take into account
nearest-neighbor (NN) interactions between magnetic ions Mn$^{3+}$($S=2$) and Mn$^{4+}$($S=3/2$) using a spin model describing a strong anisotropy on the $z$ axis.
 We have calculated the uniform and staggered magnetizations as well as the Edwards-Anderson order parameter as functions of temperature, with and without an applied magnetic field. Fitting the experimental Curie temperature at $x=0$, we estimated values of various exchange interactions in the system. The dominant one is that between Mn$^{3+}$ and Mn$^{4+}$ which is at the origin of the ferromagnetic ordering.  Effects of the very small interaction $J_2$ between NN Mn$^{3+}$ is analyzed: we show that it can cause an antiferromagnetic phase above $T_c$ which disappears at smaller $J_2$ or at Mn$^{3+}$ concentrations  smaller than 0.55. Our results show a good agreement with experiments on the magnetizations for substitution concentration $x=0.1$, 0.2 and 0.3.  We also studied the applied-field effect on the magnetization and our obtained results are compared with experiments performed at $x=10\%$.
\vspace{0.5cm}
\begin{description}
\item[PACS numbers:75.30.-m , 75.50.-y , 75.10.Hk , 05.10.Ln]
\end{description}
\end{abstract}

\pacs{PACS numbers: XXX}
\maketitle


\section{INTRODUCTION}

The study of phase transition in magnetic materials has been the subject of intensive studies both experimentally and theoretically in the last decades \cite{Zinn,DiepSP}.
In this paper, we confine ourselves to the family of perovskite compounds La$_{1-x}$A$_{x}$MnO$_{3}$ which has rich magnetic behaviors and numerous practical applications. Experiments have been
performed to determine magnetic properties of manganese oxides La$_{1-x}$A$_{x}$MnO$_{3}$, with A= Sr, Ca, Ba,.... These materials
are currently attracting a considerable attention \cite{Dagotto2001} because of the complex interplay among spins which induces a rich phase diagram as well as the colossal magnetoresistance (CMR)
phenomenon \cite{Hotta,Dagotto2000,Kim}. The rapid development of the CMR field is due mainly to its many applications in particular in spintronics \cite{Salamon,Zhu,Helmolt}.
The diagram in the phase space defined by concentration $x$, temperature $T$, magnetic field $H$ and superexchange (SE) $J$ is not quite clear yet for different compounds. Jonker and Van Santen \cite{Jonker}
have studied ferromagnetic compounds of manganese with perovskite structure. Their  properties can be understood as the result of a strong ferromagnetic exchange interaction
between nearest neighboring Mn ions via intercalated oxygen: Mn$^{3+}$-O-Mn$^{4+}$.
The double exchange (DE) mechanism developed by Zener \cite{Zener1,Zener2,Zener3} explains the existence of ferromagnetism and the metallic behavior at low temperatures. There is now a consensus to recognize that the
interesting properties observed in perovskites are fundamentally originated from the DE mechanism along the link Mn$^{3+}$-O-Mn$^{4+}$. This characteristic is at the origin of a new interesting observed phase
transition in doped manganites \cite{Hong,Bose} from a magnetically-ordered phase to the disordered phase. Recent refinement of experimental techniques and the
improvement of the sample quality have made possible to discuss critical phenomena of this transition \cite{Furukawa}. To see more details on the role of the DE, we quote a work by
Urushibara {\it et al.} \cite{Urushibara} which have investigated the transport and magnetic properties related to the insulator-metal transition of La$_{0.7}$Sr$_{0.3}$MnO$_{3}$ crystals.
 These experimental works allowed us to qualitatively understand  properties of this magnetic system. Nevertheless, these phenomena have not been quantitatively modeled, though there has been a number of works dealing with this issue from a theoretical
standpoint \cite{Hotta1,Restrepo}.


\indent In this paper we investigate by Monte Carlo (MC) simulation the magnetic properties in perovskite manganite La$_{0.7}$Sr$_{0.3}$MnO$_{3}$. Effects of Ti substitution on the
magnetic properties are studied for La$_{0.7}$Sr$_{0.3}$Mn$_{1-x}$Ti$_{x}O_{3}$ ($x$= 0.1, 0.2 and 0.25).  Recent experimental works on Ti doping \cite{Kallel,Kallel1,Kallel2,Kallel2010} show that doping with non-magnetic Ti allows one to change the relative concentration Mn$^{3+}$/Mn$^{4+}$ which in turn reduces the effective ferromagnetic interaction between them. As a consequence, one can increase the magnetic resistivity of the compound for application in spin transport.

We use a discrete spin model to express the strong Ising-like anisotropy along the $z$ axis and we take into account various types of interactions between spins in the calculation of the magnetization.  As seen in this paper, this model is justified by a good agreement with experimental measurements performed on this material.\\
\indent The paper is organized as follows: in section \ref{sectmod}, we present our model and describe the MC method. Results are reported and discussed in section \ref{sectres}, and our concluding remarks are given in
 section \ref{sectcon}.

\section {Model and Method}\label{sectmod}
\subsection{Model}

\indent We consider the simple cubic lattice with the following Hamiltonian:
\begin{equation}
 {\cal H} = -\sum_{<i,j>}J_{ij}\mathbf S_{i}\cdot \mathbf S_{j}-\mu_0H\sum_{<i>}S_{i}^z
\end{equation}
where $\mathbf S_i$ is the spin at the lattice site $i$, $\sum_{<i,j>}$ is made over spin pairs coupled through
the exchange interaction $J_{ij}$. In the following we shall take interactions between nearest-neighbors (NN) and between next nearest-neighbors (NNN) of magnetic Mn ions.   $H$ is a magnetic field applied along the $z$ axis. Let us recall that there are two kinds of Mn ions in La$_{0.7}$Sr$_{0.3}$MnO$_{3}$: Mn$^{4+}$ with spin amplitude $S=3/2$ and Mn$^{3+}$ with $S=2$. They occupy the corner sites of the simple cubic lattice. It is experimentally found that interaction between neighboring Mn$^{3+}$ and Mn$^{4+}$ is strongly ferromagnetic while that between Mn$^{4+}$ as well as that between Mn$^{3+}$ are very weakly
antiferromagnetic \cite{Kallel}.  As we will see later, to fit with experiments we need a very small interaction between NNN Mn$^{3+}$ ions. Due to the strong disorder caused by La and Sr ions, the positions of Mn$^{3+}$ and Mn$^{4+}$ are at random. In addition, when one substitutes Mn ions by non-magnetic Ti the disorder becomes even stronger due to the magnetic dilution induced by Ti substitution.  Experiments have been recently carried out on  La$_{0.7}$Sr$_{0.3}$Mn$_{1-x}$Ti$_{x}$O$_{3}$ by Kallel {\it et al.} \cite{Kallel,Kallel1,Kallel2,Kallel2010} for several $x$. We will compare our results with the data of these works.

Before defining explicitly the interactions, let us discuss about the spin model we shall use. We suppose that the spins of Mn ions lie on the $z$ axis with a strong uniaxial anisotropy.  We have first tried to calculate magnetic properties using the Heisenberg model with a strong anisotropy but we did not get an agreement with experimental data at low temperatures.  On the other hand, using a discrete Ising-like spin model, we obtain a good agreement with experiments on magnetization at various substitution concentrations in the whole temperature range as shown in the next section.

We define now the interactions. The exchange parameters $J_{ij}$ are strongly correlated to the electric structure of the compound. In 1950, Goodenough \cite{Goodenough,Goodenough1} and Kanamori \cite{Kanamori} explained
the magnetic interactions in manganites. More quantitative calculations of the magnitudes of the exchange have been attempted only recently for LaMnO$_3$ using
first-principles electronic structure methods \cite{Solovyev,Su}.
Based on the crystal and electronic structures of this system, several coupling interaction can be taken into account in the present study:\\
\indent $J_{1}$: interaction of a NN pair  Mn$^{3+}$-Mn$^{4+}$,\\
\indent $J_{2}$: interaction of  a NN pair Mn$^{3+}$-Mn$^{3+}$,\\
\indent $J_{3}$: interaction of    a NN pair  Mn$^{4+}$-Mn$^{4+}$.\\

In addition, we also introduce the following very small interactions between NNN ions:

\indent $J_{4}$: interaction between Mn$^{3+}$ NNN.\\
\indent $J_{5}$: interaction between NNN Mn$^{3+}$ and Mn$^{4+}$.\\
\indent $J_{6}$: interaction between Mn$^{4+}$ NNN.\\

 As said earlier, the DE interaction  Mn$^{3+}$-O-Mn$^{4+}$  results in a strong ferromagnetic coupling ($J_{1}$) and the SE interactions Mn$^{3+}$-O-Mn$^{3+}$ ($J_{2}$) and
Mn$^{4+}$-O-Mn$^{4+}$ ($J_{3}$) give rise to very weak antiferromagnetic exchange integrals \cite{Jonker,Birsan,Anderson}.  We shall see below that though very small, these antiferromagnetic interactions cannot be neglected: they are at the origin of magnetic behaviors at very low $T$ and of the antiferromagnetic phase in a small temperature region above $T_c$, as discussed in the next section.  As for $J_4$, $J_5$ and $J_6$ they are assumed to be even smaller than $J_2$ and $J_3$. They are to be used for fine tuning of the fit. We have tested them while fitting with Ti substitution shown below. Only $J_4$ may be necessary since the high Mn$^{3+}$ concentration (70\%) allows for such a coupling to be visible. So from now on, we will neglect $J_5$ and $J_6$ for clarity.

\subsection{Method}
 We have conducted standard MC simulation on  samples of dimension $N=L\times L\times L$, where $L$ is the number of simple cubic cells in each of the three Cartesian directions. Periodic boundary conditions are used in all directions.
 Simulations have been carried out for different lattice sizes ranging from $12^3$ to $20^3$ lattice cells.  Some runs with $N=24^3$, 28$^3$ and 32$^3$ have also been performed  to check finite-size effects as will be shown below.

The procedure of our simulation can be split into two steps. The first step consists in equilibrating the lattice at a given temperature.
For the thermalization, two difficult regions are the low-temperature ($T$) one and the critical region. The low-$T$ region have small 'update' probabilities [proportional to exp(-E/T)] and therefore needs longer runs to ensure the equilibrium state before averaging: a simple check of energy behavior with time evolution and temperature evolution at low $T$ suffices. Near the transition temperature $T_c$, the critical slowing-down also necessitates long runs to ensure a good statistical average. Test runs with different run lengths were performed to estimate necessary run length before real simulations.
The second step, when equilibrium is reached, we determine thermodynamic properties by taking thermal averages of various physical quantities \cite{DiepTM,Metropolis}. Starting from a random spin configuration as the initial condition for the MC simulation, we have calculated
the internal energy per spin $E$, the specific heat $C_V$, the magnetic susceptibility $\large {X}$, the magnetization of each sublattice and the total magnetization, as functions of temperature $T$ and magnetic field $H$.
The MC run time for equilibrating is about $10^5$ MC steps per spin. The averaging is taken, after equilibrating, over $10^5$ MC steps.

Since the system has a strong disorder (random mixing of Mn ions), we will see that the size effects are not significant from $20^3$ sizes. Most of the simulations have been therefore carried out at this size using 20 to 30 samples for largest $x$. For smaller $x$ and large sample sizes the configuration average needs smaller number of configurations (about 10 to 20). The results on $T_c$  vary over an interval of 4K with various disorder samples.  So errors are $\pm 2$ K around the mean value. These errors cover smaller errors $\pm 1$ K due to the peak determination of $T_c$.

The total magnetization $M_t$ is defined by
\begin{equation}
M_t=\frac{1}{N}\langle\sum_{i}\mathbf S_i\rangle\label{mtot}
\end{equation}
where the sum is performed over spins of both Mn$^{3+}$ and Mn$^{4+}$ and $\langle...\rangle$ indicates the statistical time average.  The magnetizations of Mn$^{3+}$ and Mn$^{4+}$ ($M_1$ and $M_2$), their staggered magnetizations ($M_{s1}$ and $M_{s2}$) and the Edwards-Anderson order parameter $Q_{EA}$ are defined by
\begin{eqnarray}
M_{\ell}&=&\frac{1}{N_{\ell}}\langle\sum_{i}\mathbf S_i\rangle \label{msub}\\
M_{s\ell}&=&\frac{1}{N_{\ell}}\langle\sum_{i}(-1)^i\mathbf S_i\rangle \label{msta}\\
Q_{EA}(\ell)&=&\frac{1}{N_{\ell}}\sum_{i}\langle \mathbf S_i\rangle^2 \label{qeas}
\end{eqnarray}
where the sum is taken over Mn$^{3+}$ ($\ell=1$) or Mn$^{4+}$ ($\ell=2$) with $N_{\ell}$ being the number of spins of each kind. Note that $Q_{EA}$ is calculated by first taking the time average of each spin and secondly taking the spatial average over all spins. This parameter is used to calculate the freezing degree of the spins when a long-range ordering is absent or the nature of ordering is unknown such as in spin glasses or in disordered systems \cite{Ngo2014,EdwardsAnderson,BinderYoung,Mezard}.

For La$_{0.7}$Sr$_{0.3}$MnO$_{3}$ magnetic ions Mn$^{3+}$ and Mn$^{4+}$ have concentrations 0.70 and 0.30, respectively \cite{Kallel}.  Hereafter, we study this ``mother" compound and the case where Ti ions replace a fraction $x$ of Mn$^{4+}$.
Note that in our procedure, we first generate the pure state (Mn$^{3+}$ only) which is the fully antiferromagnetic state which has the site 'parity'. Then we replace Mn$^{3+}$ by Mn$^{4+}$ at randomly chosen sites with a concentration of 0.3. In doing so we conserve the site 'parity' because we do not exceed the percolation limit for the simple cubic structure. This is the reason why we define the staggered magnetization which is useful for detecting an antiferromagnetic phase. Of course, this is artificially created state but physically the results will not change if we have large domains of Mn$^{3+}$ in the compound. The substitution of Mn$^{4+}$ by Ti is next performed also at randomly chosen Mn$^{4+}$ ions.

\section{Results}\label{sectres}
\subsection{Properties of La$_{0.7}$Sr$_{0.3}$MnO$_{3}$}
Let us examine first the case without Ti doping in order to have an idea about the role of each interaction $J_1$, $J_2$, $J_3$ and $J_4$.

As said earlier, experiments found that $J_1$ dominates and gives rise to the ferromagnetic ordering up to very high temperatures $T_c=369$ K \cite{Kallel}.   For $H=0$, we have fitted the MC transition temperature with experimental value $T_c=369$ K using the mean-field approximation:
\begin{equation}\label{MFTC}
T_c=\frac{2}{3k_B}ZS(S+1)J_{eff}
\end{equation}
where $Z=6$ is coordination number (all occupied neighbors) and $S=\sqrt{3}$  the effective spin value calculated from $S^2=S(\mbox{Mn}^{4+})S(\mbox{Mn}^{3+})=2\frac{3}{2}=3$. Putting $T_c=369$ K, we obtain $J_{eff}\simeq 25.1 $ K.  Note that in magnetic materials with Curie temperatures at room or higher temperatures the effective exchange interaction is of the order of several dozens of Kelvin \cite{Magnin2012}, just as what we found here. It is not easy to determine each of the exchange interactions defined earlier. Fortunately, we know that $J_1$ is much larger than the other interactions \cite{Kallel}. Writing
in the mean-field spirit $J_{eff}\simeq J_1+|J_2|+|J_3|+J_4$ and taking, after comparing our calculated magnetizations with several trying values of $J_2$, $J_3$ and $J_4$ we find the best fits of magnetizations obtained with $J_2=J_3=J_4=-J_1/120$, as shown below. The best estimated values of the main exchange interactions are $J_1\simeq 24.5$ K and $J_2=J_3=-J_4\simeq -0.20$ K.  We will take $J_3=-J_4=-0.20$ K for all calculations and we will discuss below the effects of $J_2$ since this parameter is found to be more relevant than the other two.

\begin{figure}[ht!]
\centering
\includegraphics[width=7cm,angle=0]{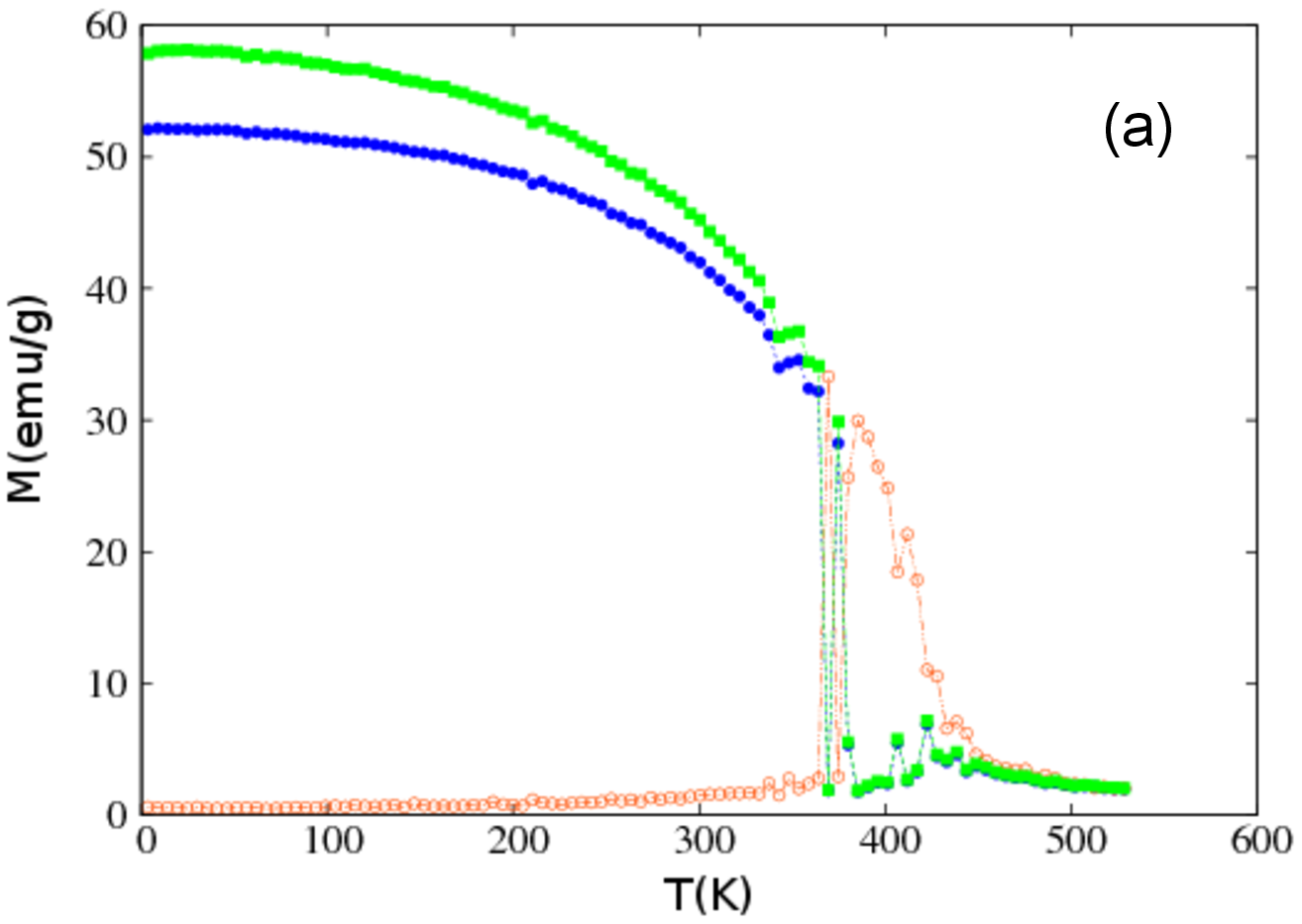}
\includegraphics[width=7cm,angle=0]{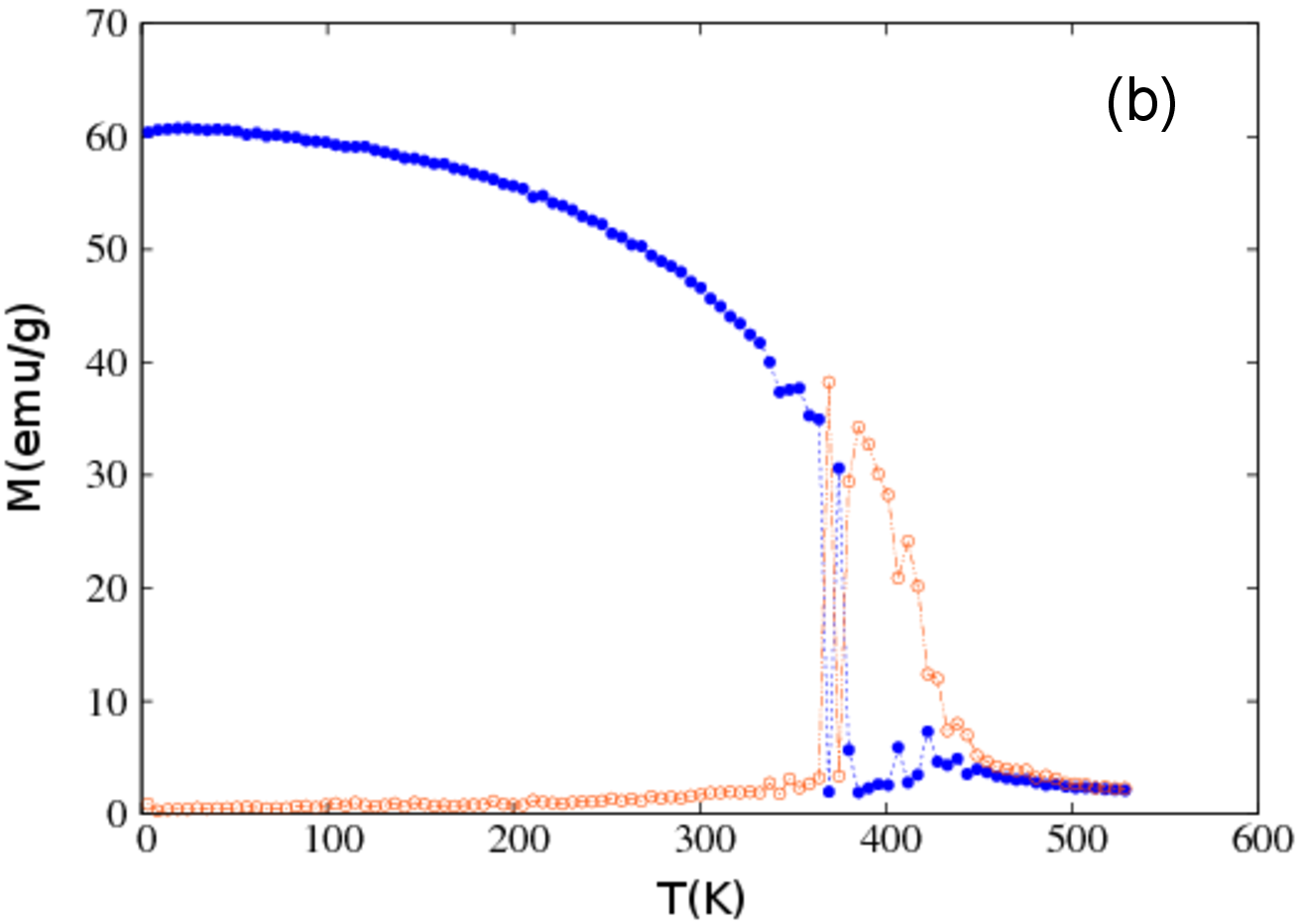}
\includegraphics[width=7cm,angle=0]{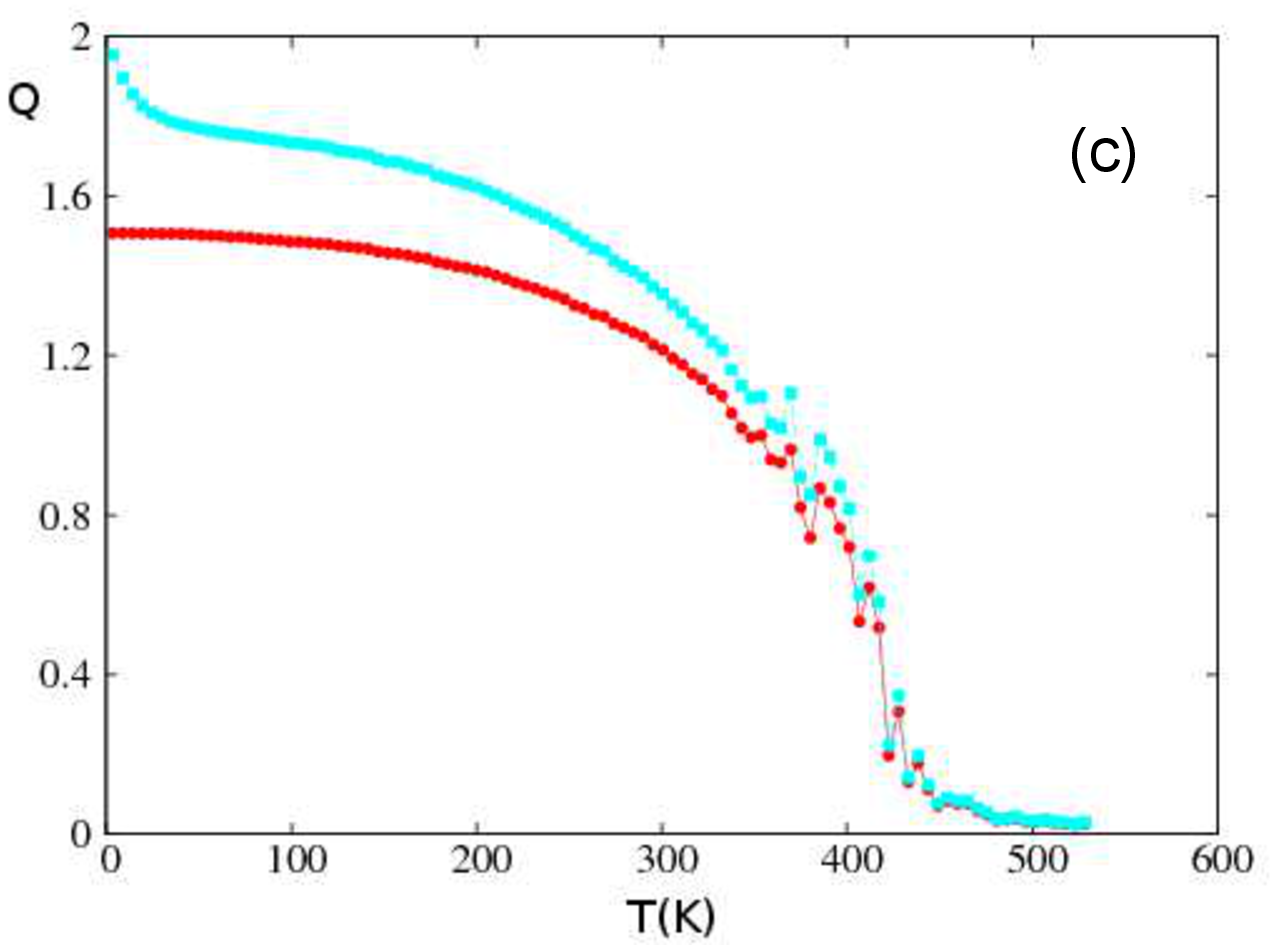}
\caption{(Color online) La$_{0.7}$Sr$_{0.3}$Mn$_{0.7}^{3+}$Mn$_{0.3}^{4+}$O$_{3}$ (no Ti doping): Results of Monte Carlo simulation for $J_1=24.5$ K, $J_2=J_3=-J_4=-0.20$ K: (a) the magnetization (blue circles) and staggered magnetization (orange void circles) of  the subsystem of Mn$^{4+}$, (b) those of the subsystem of Mn$^{3+}$,
(c) Edwards-Anderson order parameter $Q_{EA}$ versus temperature $T$ for Mn$^{4+}$ (red circles) and Mn$^{3+}$ (cyan squares). The total magnetization is also shown in the top figure by green squares. \label{figms}}
\end{figure}

We show in Fig. \ref{figms} the total magnetization $M_t$ (green squares) and the sublattice magnetizations $M_{1,2}$ and sublattice staggered magnetizations $M_{s1,s2}$.  We calculated the magnetization in the simulation using parameters representing the magnetic moments of Mn$^{3+}$ and Mn$^{4+}$. We next fit our values found with simulations at the lowest $T$ with values found experimentally (see for example experimental data in the references \cite{Kallel,Kallel1,Kallel2,Kallel2010}).  The unit used for magnetizations in this paper (emu/g) is by this fit.  Several remarks on Fig. \ref{figms} are in order:

(i) $M_t$ shows a transition at $T_c=369$ K

(ii) The magnetizations of Mn$^{3+}$  and Mn$^{4+}$ vanish rather abruptly at that temperature. This suggests that the transition may be weakly of first order.

(iii) Surprisingly, the staggered magnetizations become non zero for $T>T_c$ up to $\simeq 427$ K. This indicates that
 an antiferromagnetic ordering appears above the ferromagnetic phase.

(iv) The Edwards-Anderson order parameters show  ``strong fluctuations" at and above $T\simeq 369$ K.  The fact that $Q_{EA}$ is not zero between 369 K and 427 K means that there is a phrase which is not paramagnetic. However $Q_{EA}$ alone cannot determine the nature of the ordering. The non-zero staggered magnetization in this temperature zone helps confirm an antiferromagnetic phase.

(v) At very low $T$, the antiferromagnetic interaction $J_2$ between Mn$^{3+}$ affects its magnetization behavior: weak values (smaller than -0.20 K) keep the ferromagnetic ordering near $T=0$ [Fig. \ref{figms}(b)] but induces a reduction of $Q_{EA}$ in Fig. \ref{figms}(c). This reduction is a signature of some dynamical motion of Mn$^{3+}$. We return to this point later while showing results with other values of $J_2$. Note that $J_2$ does not affect the magnetization and $Q_{EA}$ of Mn$^{4+}$.

\begin{figure}[ht!]
\centering
\includegraphics[width=7cm,angle=0]{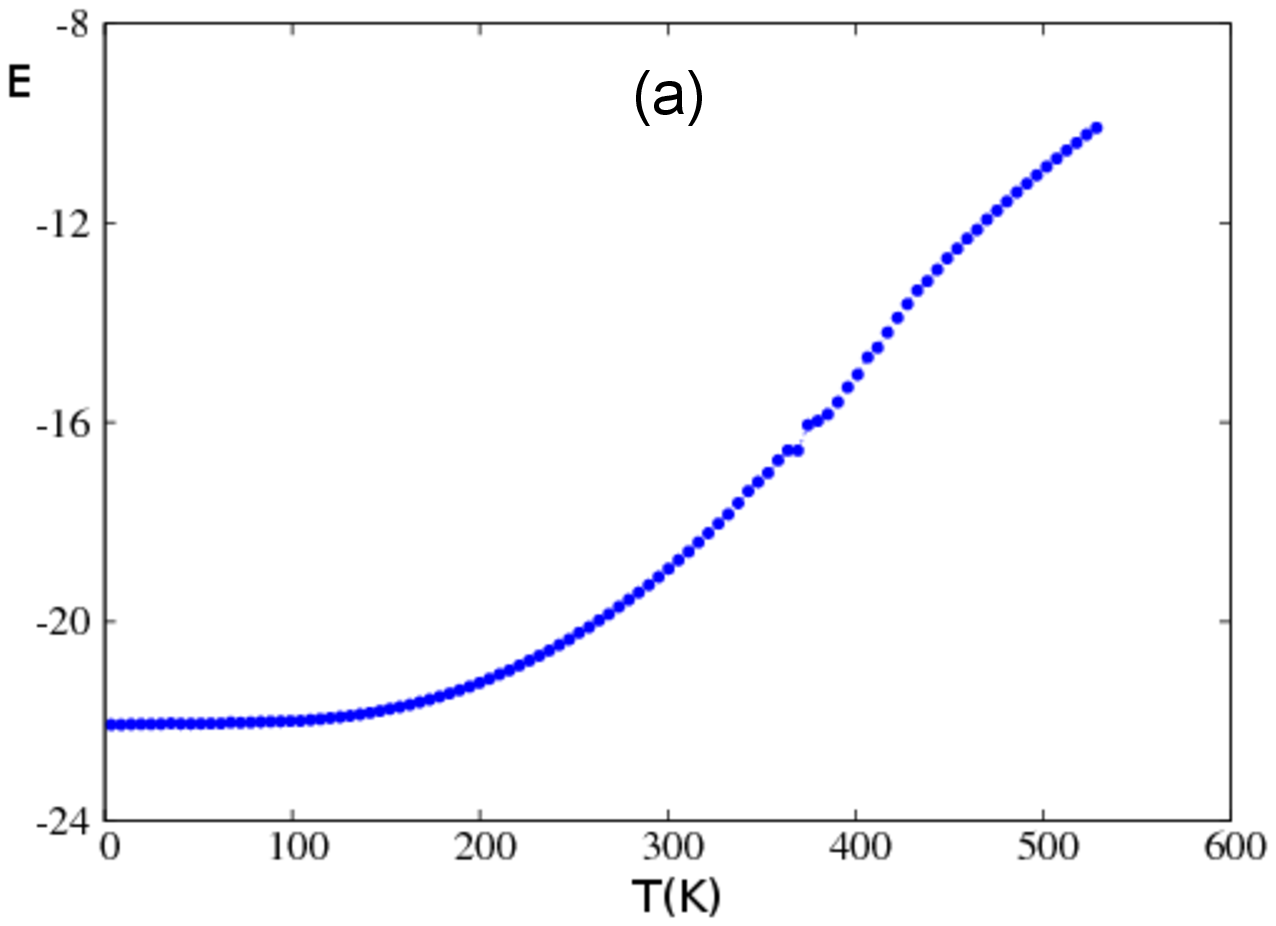}
\includegraphics[width=7cm,angle=0]{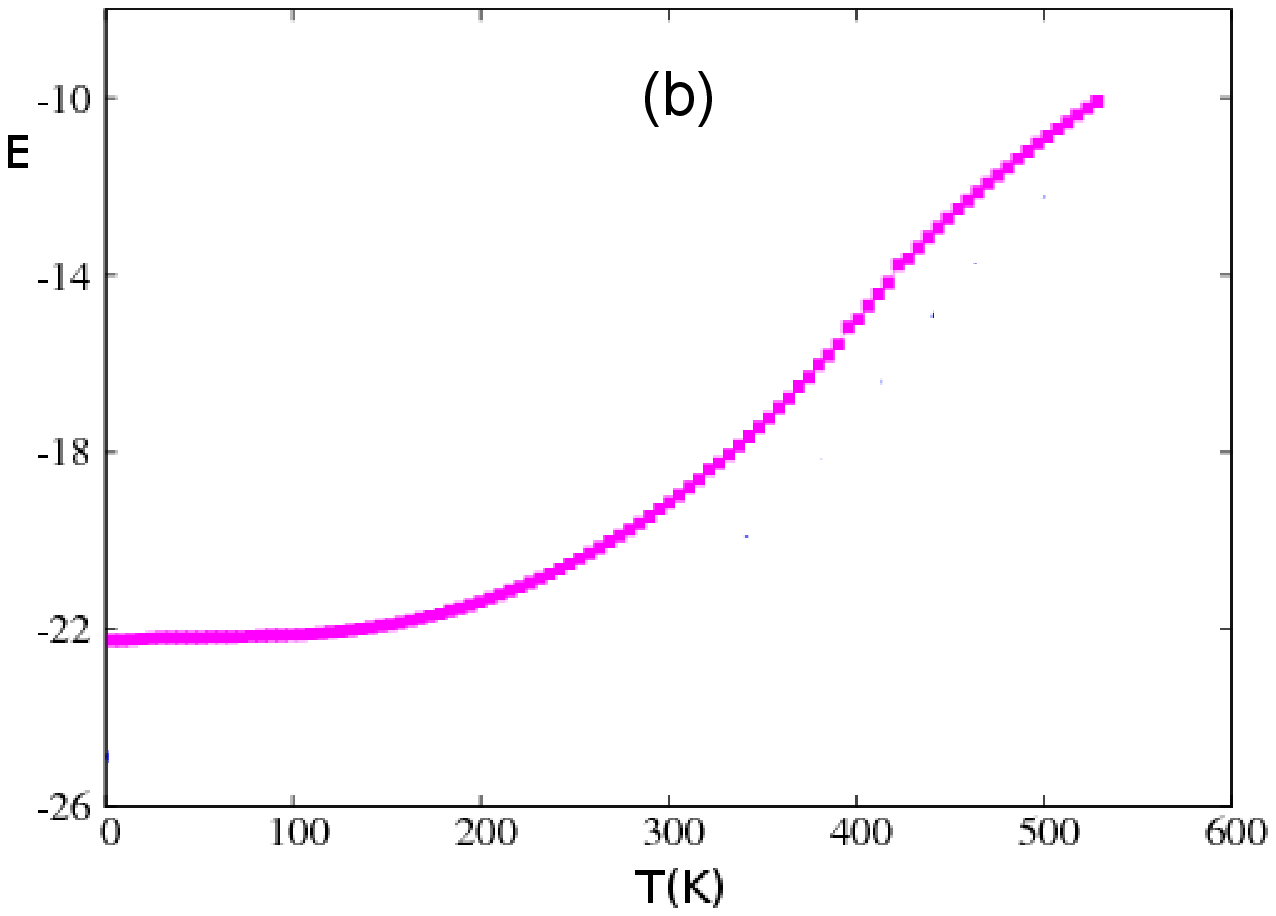}
\caption{(Color online) La$_{0.7}$Sr$_{0.3}$Mn$_{0.7}^{3+}$Mn$_{0.3}^{4+}$O$_{3}$: Internal energy $E$ per spin, in unit of $10^{-3}$ eV, versus $T$ with $J_1=24.5$ K and $J_3=-J_4=-0.20$ K for two values of $J_2$: (a) $J_2=-0.20$ K where the phase transition occurs at $T\simeq 369$ K with a jump indicating a first-order character, (b) $J_2=-0.08$ K (blue circles) where the phase transition is seen at $T\simeq 422$ K. See text for comments.\label{fige}}
\end{figure}

To show that the phase transition at $T_c=369$ K is of first-order, we show in Fig. \ref{fige}(a) the energy versus $T$ using the same parameters as those of Fig. \ref{figms}, namely $J_1=24.5$ K, $J_2=J_3=-J_4=-0.20$ K.  We observe a discontinuity at 369 K. For comparison, we show in Fig. \ref{fige}(b) the case where $J_2$ is smaller, $J_2=-0.08$ K. We observe that the transition temperature is very sensitive to $J_2$: it moves to $\simeq$ 422 K.

We have made several simulations to search for the parameter(s) responsible for the antiferromagnetic phase: we find that the value of $J_3$ does not affect neither the antiferromagnetic phase nor the value of $T_c$ as long as it is of the order of $\simeq -0.20$ K. This is easily understood because the concentration of Mn$^{4+}$ is small (30\%) so that the number of NN pairs of Mn$^{4+}$ with interaction $J_3$ not intercalated by  Mn$^{3+}$ is negligible.

\begin{figure}[ht!]
\centering
\includegraphics[width=7cm,angle=0]{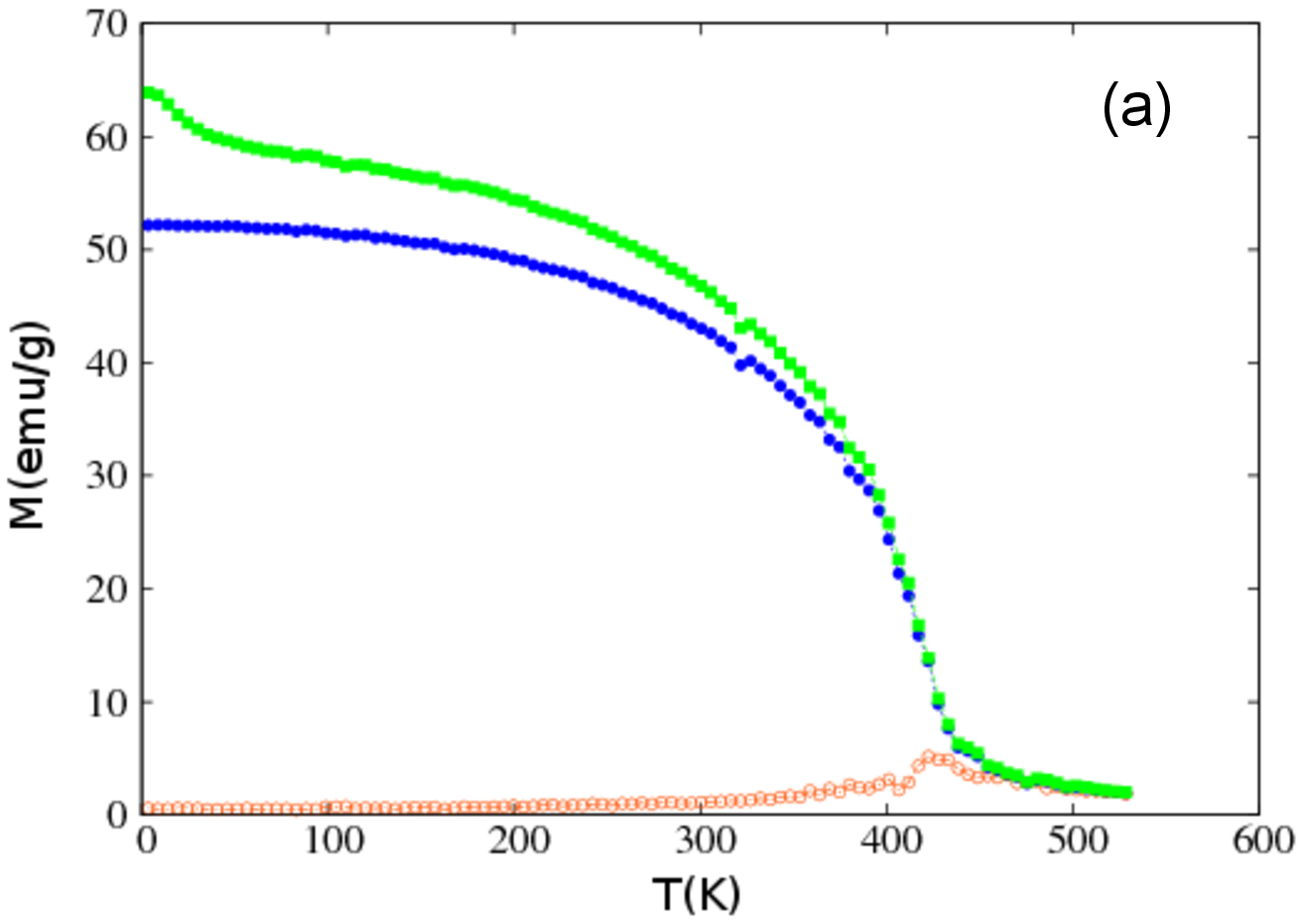}
\includegraphics[width=7cm,angle=0]{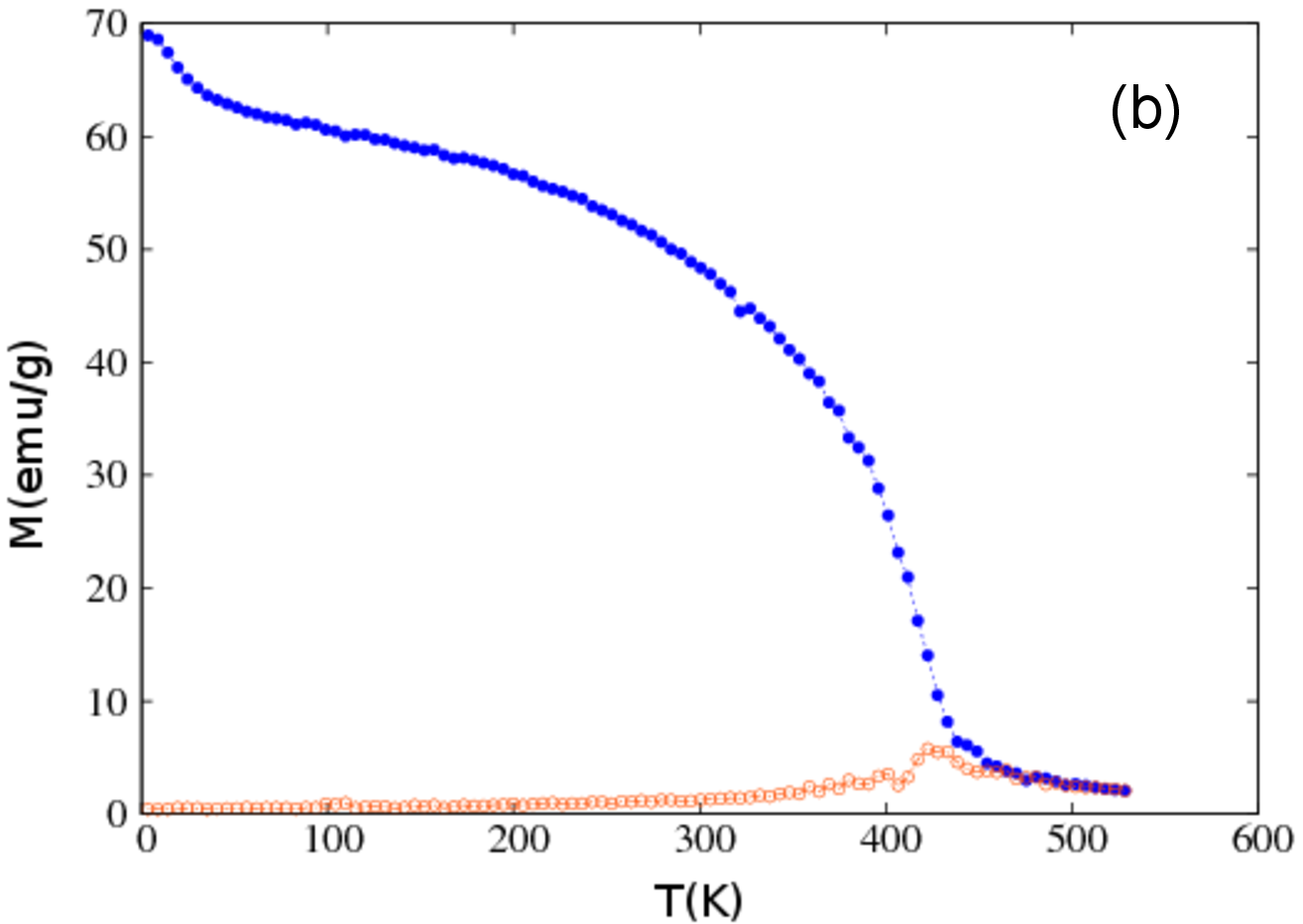}
\caption{(Color online) La$_{0.7}$Sr$_{0.3}$Mn$_{0.7}^{3+}$Mn$_{0.3}^{4+}$O$_{3}$: Effect of very small $J_2$: (a) magnetization (blue circles) and staggered magnetization (orange void circles) of Mn$^{4+}$. The total magnetization averaged over the whole system is also shown by green squares. (b) magnetization (blue circles) and staggered magnetization (orange void circles) of Mn$^{3+}$. Used values: $J_2=-0.08$ K ($J_1=24.5$ K, $J_3=-J_4=-0.20$ K).
The antiferromagnetic phase seen with $J_2=-0.20$ K shown in Fig. \ref{figms} disappears with this small value of $J_2$.
\label{figms002}}
\end{figure}

 We have checked that the antiferromagnetic phase disappears if the antiferromagnetic interaction $J_2$ between Mn$^{3+}$ is small.  Figure \ref{figms002} shows no antiferromagnetic phase for $J_2\simeq -0.08$ K.   This is not surprising because the antiferromagnetic order does not survive at high $T$ with such small $J_2$.

\begin{figure}[ht!]
\centering
\includegraphics[width=7cm,angle=0]{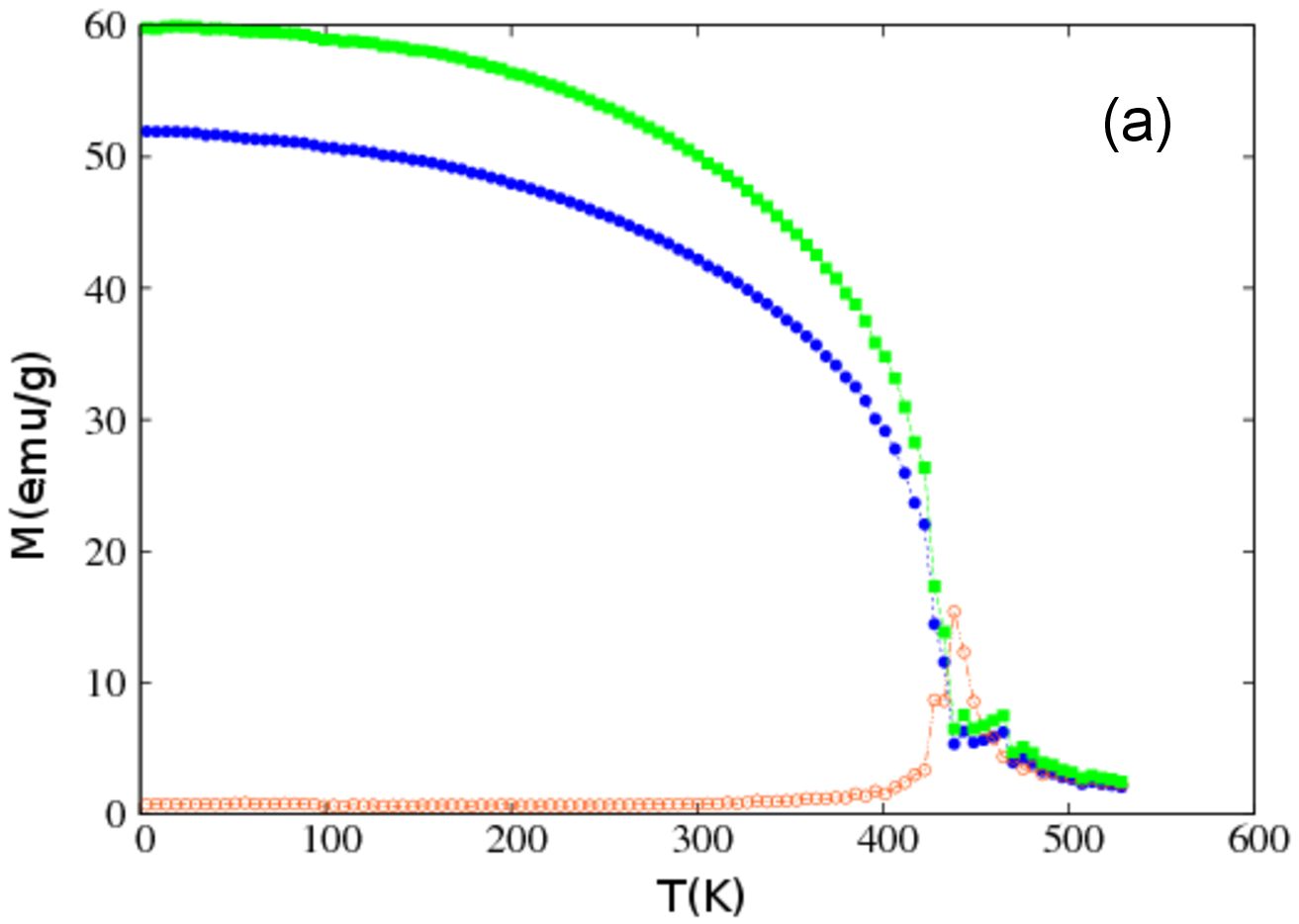}
\includegraphics[width=7cm,angle=0]{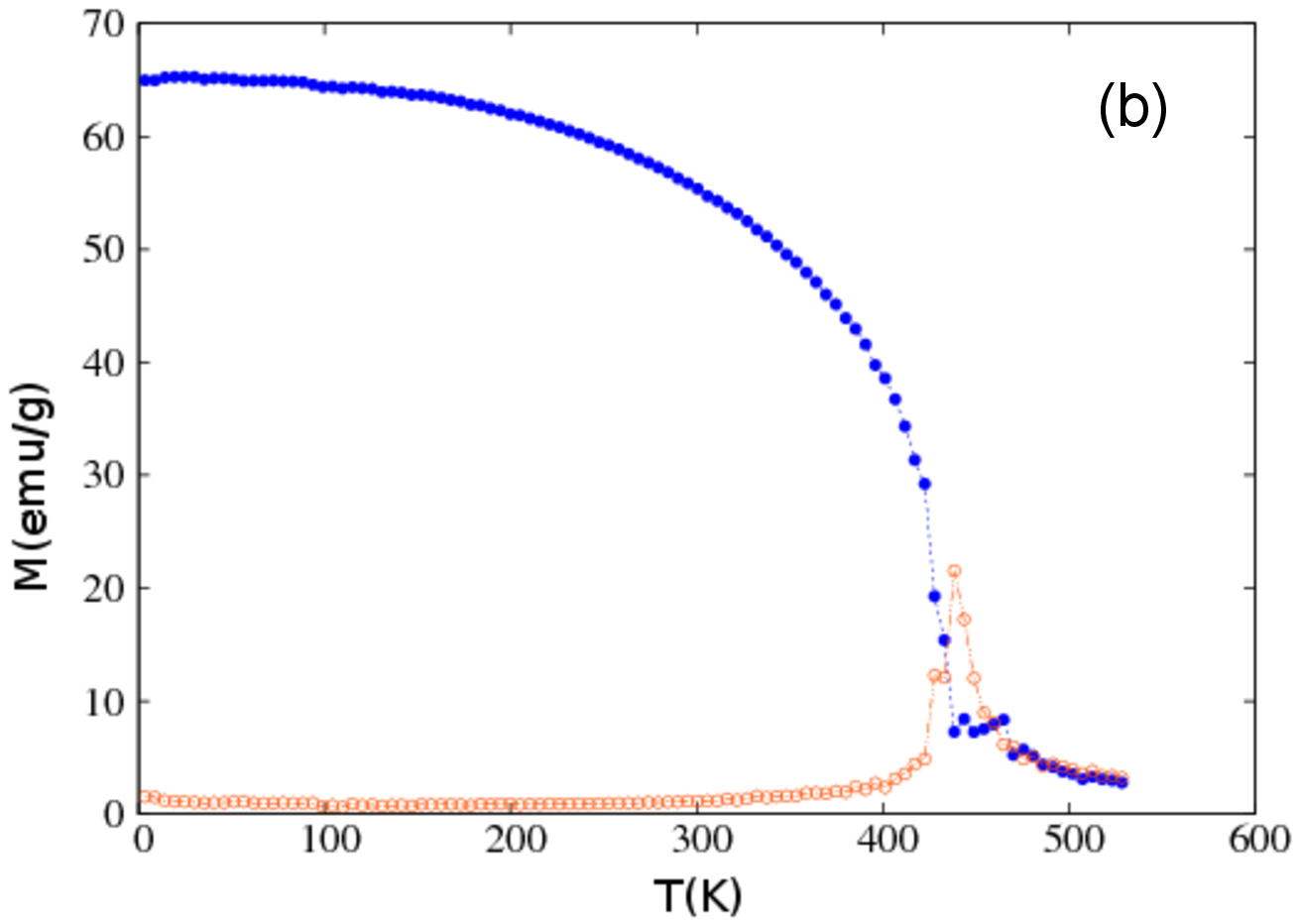}
\caption{(Color online) La$_{0.6}$Sr$_{0.4}$Mn$_{0.6}^{3+}$Mn$_{0.4}^{4+}$O$_{3}$: Results at 60\% of Mn$^{3+}$ (a)  magnetization (blue circles) and staggered magnetization (orange void circles) of  Mn$^{4+}$,  (b) those of Mn$^{3+}$. We have used $J_2=-0.20$ K ($J_1=24.5$ K, $J_3=-J_4=-0.20$ K).
The total magnetization averaged over the whole system is also shown by green squares in the top figure. The antiferromagnetic phase is reduced with respect to that at 70\% shown in Fig. \ref{figms}.
\label{figmsc06}}
\end{figure}

The disappearance of the antiferromagnetic phase can come also from the Mn$^{3+}$ concentration: by symmetry argument, if the concentrations of Mn$^{3+}$ and Mn$^{4+}$ are 50\% each, then each Mn$^{3+}$ is surrounded in most cases by Mn$^{4+}$ and vice-versa. As a consequence, there are no rooms for a sufficient number of antiferromagnetic pairs of Mn$^{3+}$-Mn$^{3+}$ and Mn$^{4+}$-Mn$^{4+}$. This prevents an overall antiferromagnetic ordering to occur. When the concentrations are not symmetric such as 70\%-30\% there are always domains of the majority spins which are antiferromagnetically coupled, leading to antiferromagnetic phase above the ferromagnetic phase. This antiferromagnetic phase, if it exists in La$_{0.7}$Sr$_{0.3}$MnO$_{3}$, has not been experimentally observed.
An example with 60\% Mn$^{3+}$ is shown in Fig. \ref{figmsc06}: the antiferromagnetic phase exists but in a region narrower than that for 70\%. We have verified that if the concentration of Mn$^{3+}$ becomes smaller than 0.55, there is no more antiferromagnetic phase above the ferromagnetic phase.
At this stage, it is worth noting that in spin glasses that weak disorder can cause a second phase above the spin-glass phase. An example is the $\pm J$ Ising spin glass: if the percentage of $-J$ bonds is large with respect to that of $+J$ then the antiferromagnetic phase exists above the spin-glass phase in a region called reentrance \cite{Ngo2014,BinderYoung}.  We will take $J_2=-0.20$ K for the results shown below with Ti substitutions at $x=0.1$, 0.2 and 0.25.

\begin{figure}[t!]
\centering
\includegraphics[width=7cm,angle=0]{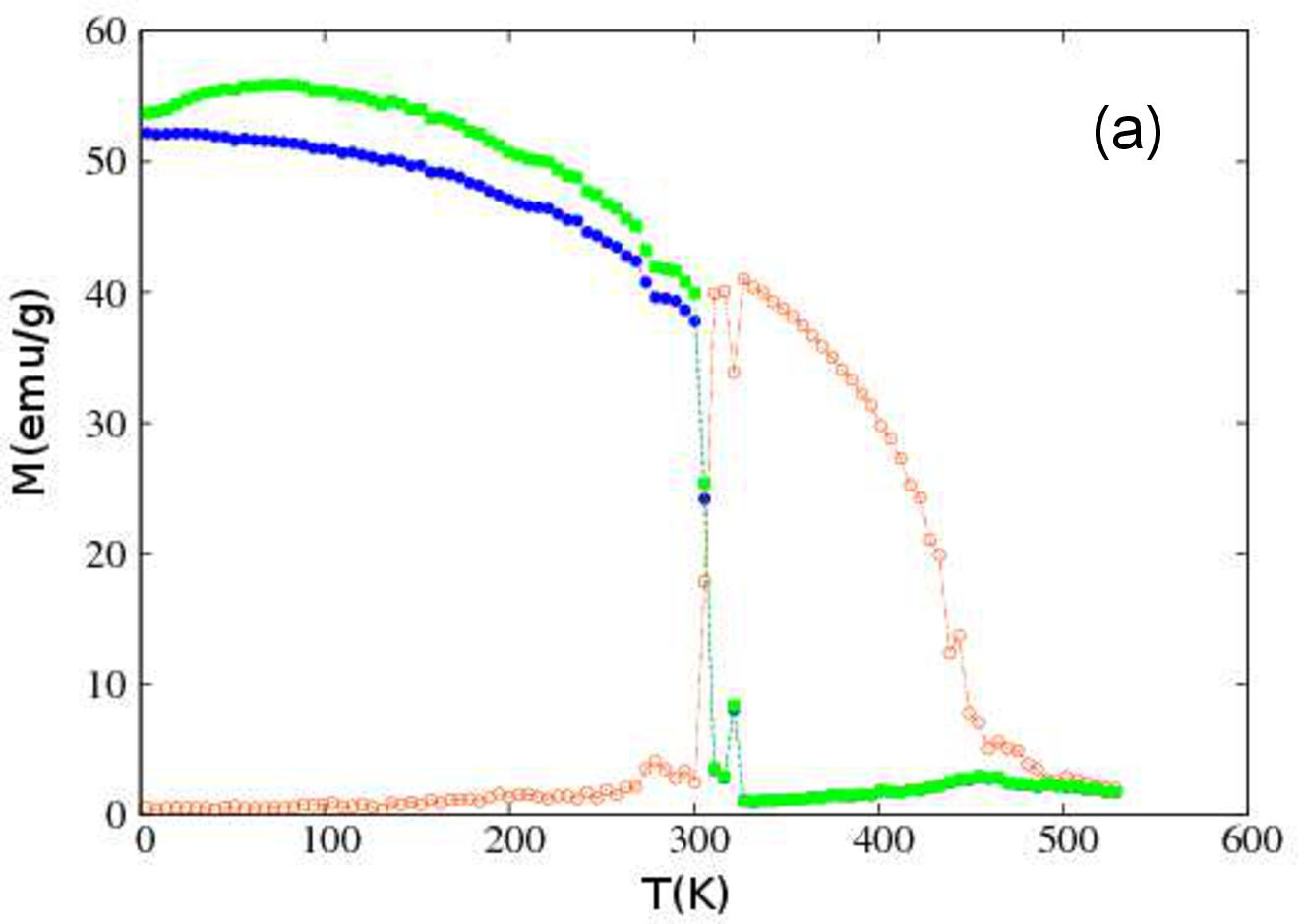}
\includegraphics[width=7cm,angle=0]{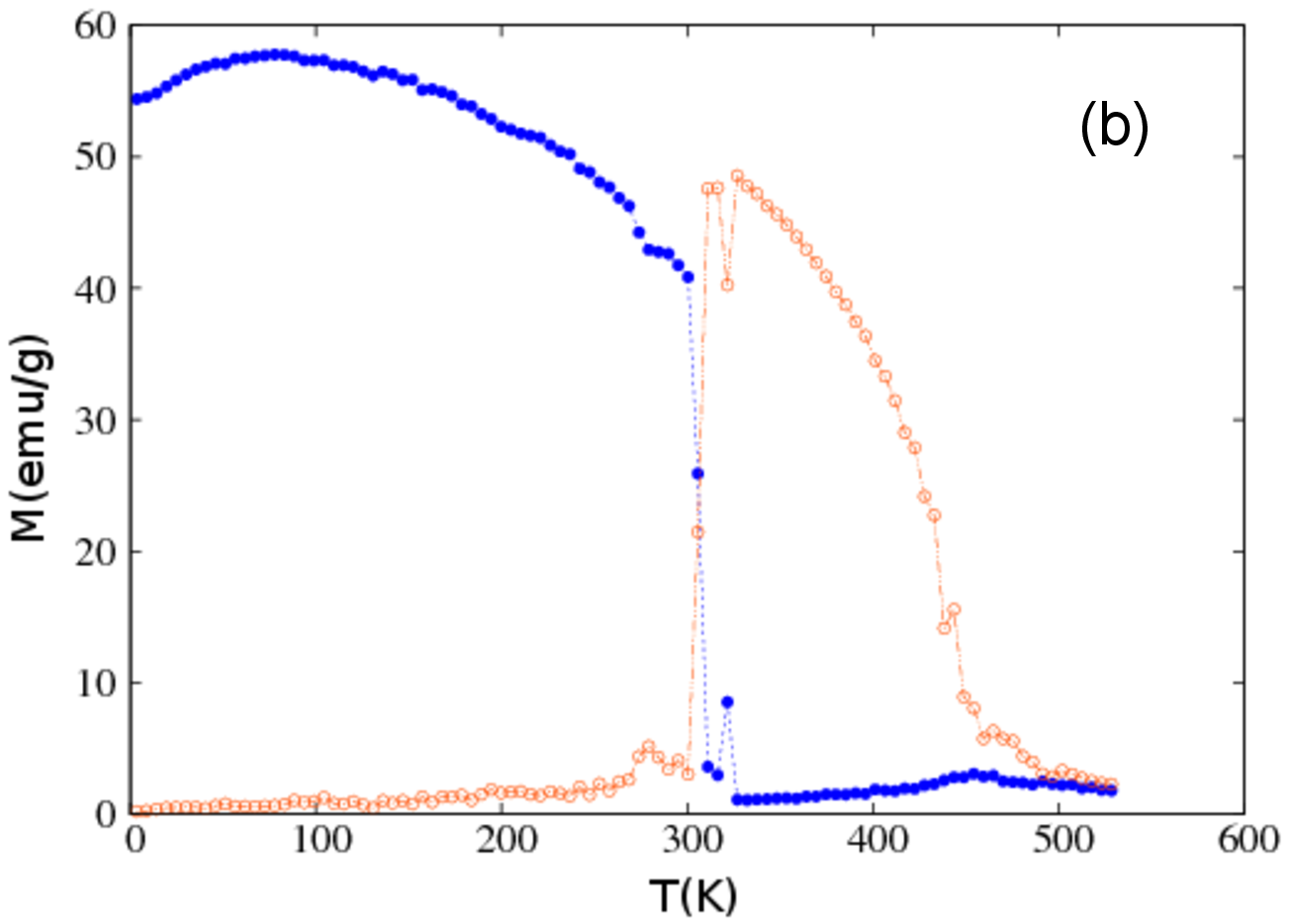}
\includegraphics[width=7cm,angle=0]{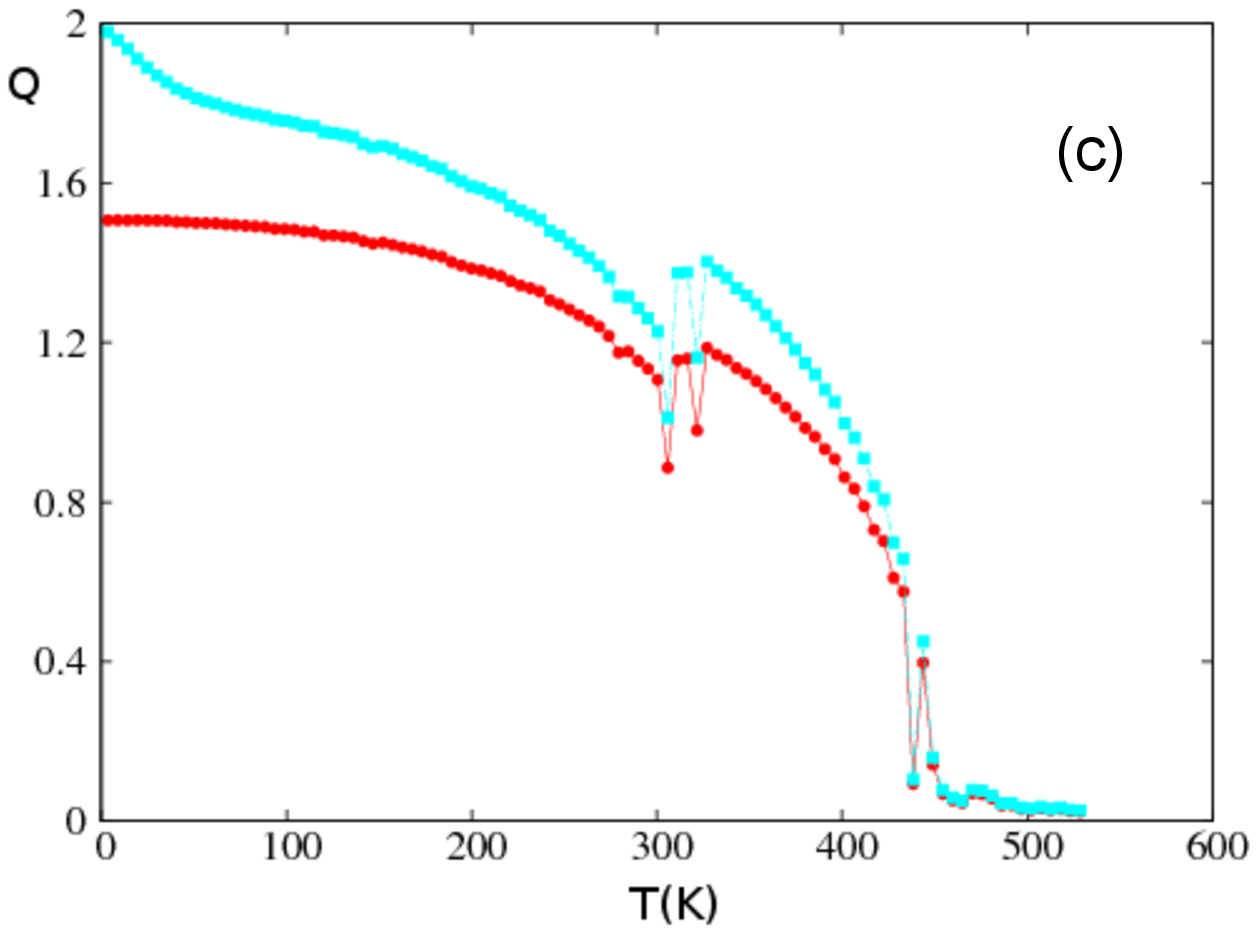}
\caption{(Color online) La$_{0.7}$Sr$_{0.3}$Mn$_{0.7}^{3+}$Mn$_{0.3}^{4+}$O$_{3}$: Effect of a large value of $J_2$: $J_2=-0.40$ K ($J_1=24.5$ K, $J_3=-J_4=-0.20$ K) (a) magnetization (blue circles) and staggered magnetization (orange void circles) of Mn$^{4+}$, (b) those of Mn$^{3+}$, with , (c) Edwards-Anderson order parameters for Mn$^{4+}$ (red circles) and Mn$^{3+}$ (cyan squares).
The total magnetization averaged over the whole system is also shown by green squares in the top figure.
The larger value of $J_2$ widens the antiferromagnetic phase with respect to that at 70\% shown in Fig. \ref{figms}. See text for other comments.
\label{figms01}}
\end{figure}

Figure \ref{figms01} shows the results simulated with a large value of $J_2$, namely -0.40 K, twice larger than the value used in Fig. \ref{figms}. As expected, the abrupt character of the transition increases and the antiferromagnetic phase becomes stronger in a larger temperature range.  Note the strong variation of $Q_{EA}$ at the transition from ferromagnetic to antiferromagnetic phase.  Furthermore, at this strong value of $J_2$, there are at low $T$ a number of antiferromagnetic NNN pairs which 'resist' the strong ferromagnetic Mn$^{3+}$-Mn$^{4+}$ coupling. This yield a reduction of the Mn$^{3+}$ magnetization at low $T$ ($<$ 100 K) as seen in Fig. \ref{figms01}(b). As already commented in remark (v) on Fig. \ref{figms}, this instability is signaled by a reduction of $Q_{EA}$ in this low-$T$ region.  By comparing the values at low $T$ for three cases with increasing $J_2$: $M=69$ emu/g for $J_2=-0.08$ K [Fig. \ref{figms002}(b)], $M=60$ emu/g for $J_2=-0.20$ K [Fig. \ref{figms}(b)] and $M=54$ emu/g for $J_2=-0.40$ K [Fig. \ref{figms01}(b)], we conclude that even the ferromagnetic ordering dominates the compound at low $T$, the stronger $J_2$ yields a smaller magnetization of Mn$^{3+}$. However, this reduction of magnetization at low $T$ is weakened at higher $T$: the curvature of $M$ shows a decreasing tendency with increasing $T$ for $T>100$ K.

\begin{figure}[ht!]
\centering
\includegraphics[width=6.0cm,angle=0]{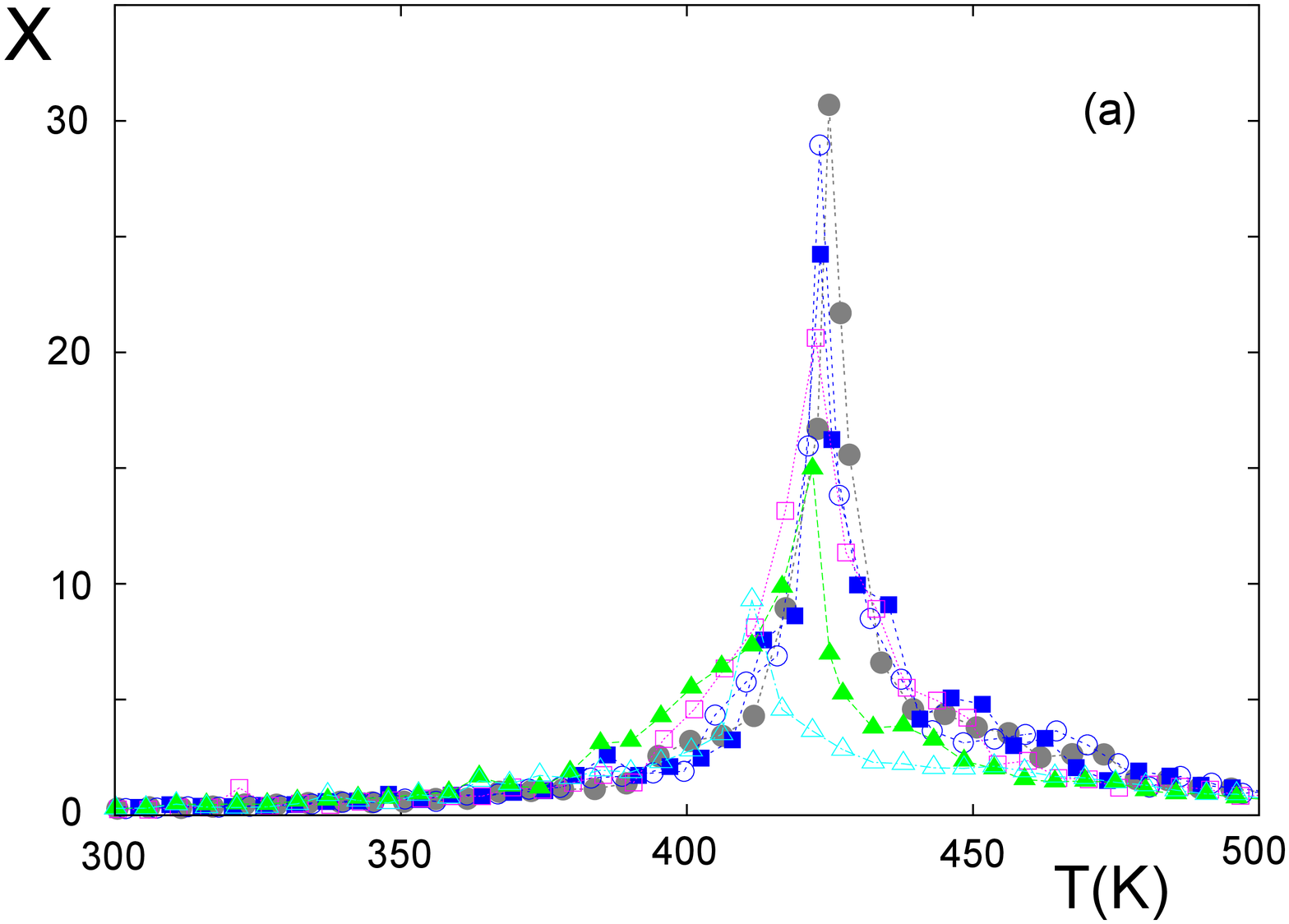}
\includegraphics[width=6.5cm,angle=0]{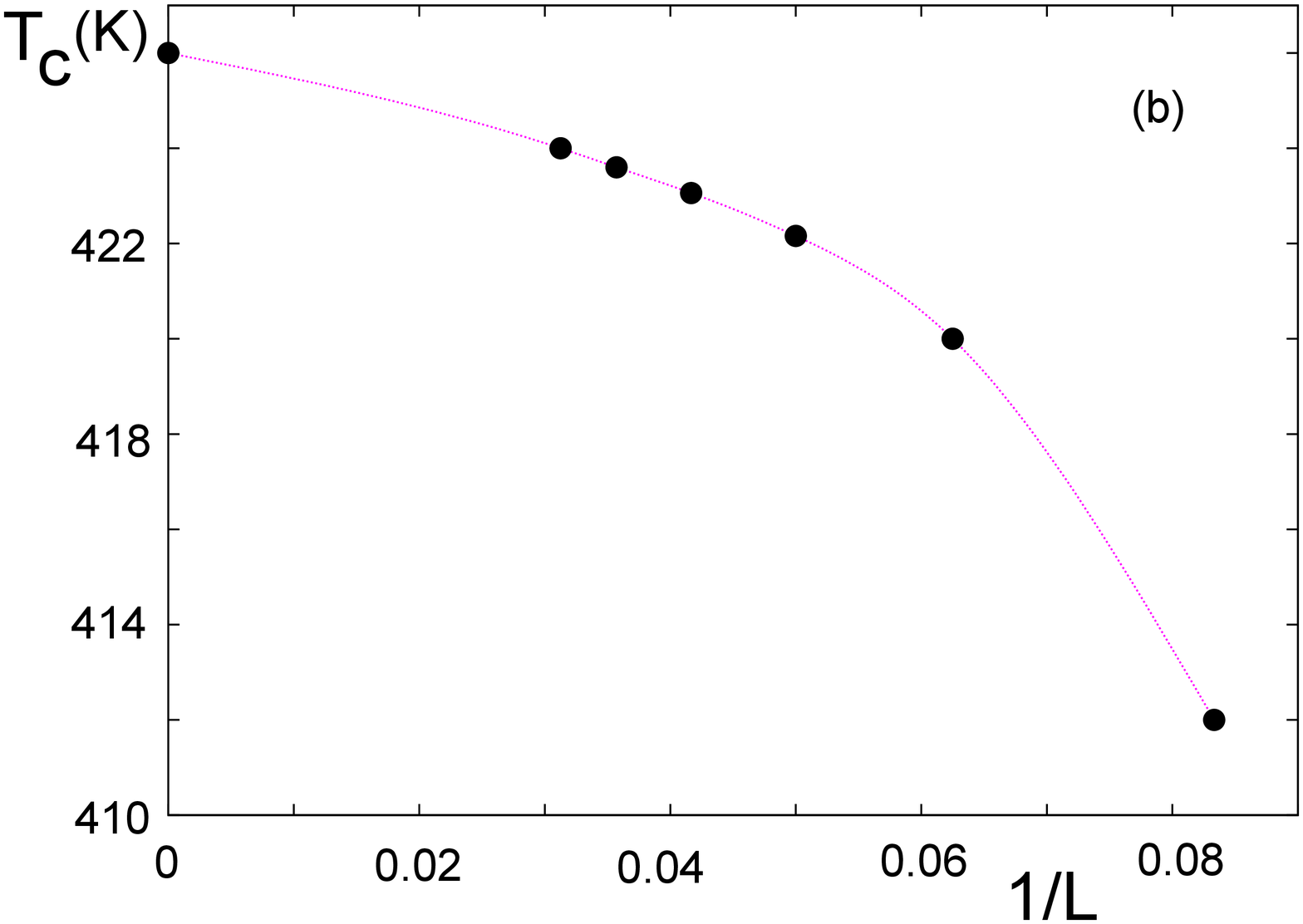}
\caption{(Color online) La$_{0.7}$Sr$_{0.3}$Mn$_{0.7}^{3+}$Mn$_{0.3}^{4+}$O$_{3}$: Example of finite-size effects for $J_1=24.5$ K, $J_3=-J_4=-0.20$ K and $J_2=-0.08$ K, (a) Susceptibility $\large {X}$ calculated from fluctuations of the total magnetization for sizes $N=L^3$ with $L$=12 (cyan void triangles), 16 (green filled triangles), 20 (red void squares), 24 (blue filled squares), 28 (blue void circles) and 32 (gray filled circles), many points have been removed for the sake of clarity, (b) Curie temperature versus $1/L$, note that the point for $1/L=0$ (infinite lattice size) is extrapolated using the smooth function of csplines. \label{figsize}}
\end{figure}

Let us in Fig. \ref{figsize} how we determine $T_c$ in order to establish a phase diagram in the parameter space ($T_c,P)$ where $P$ is the concentration of Mn$^{3+}$.  As said in the previous section, we have carried out most simulations with the lattice size $20^3$ because from this size up, finite-size effects are not strong. There are two error estimations: the first one consists of making many runs in the critical region using temperature steps as small as 2 K. The peak of the susceptibility is identified with error $\pm 1$ K. The second error estimation consists of making runs with differently generated disordered samples: we observed that $T_c$ changes in an interval of $4$ K. So the error over disorder is $\pm 2$ K around the median value. Fortunately, this error includes the error due to discrete temperature steps.  We show in Fig. \ref{figsize} an example of finite-size effects for $J_1=24.5$ K, $J_3=-J_4=-0.20$ K and $J_2=-0.08$ K. This case is an easy one due to the absence of the antiferromagnetic case.   The sizes used are $L=12$, 16, 20, 24, 28 and 32.  In spite of small $J_2$, the susceptibility has strong fluctuations around each peak. The peak positions can be recognized without difficulty. For curve presentation we used the most typical runs, namely the ones which correspond to the mean values of $T_c$  at each size. The error bars around each point are $\pm 1$. The extrapolated point at the infinite size is $426\pm 1$ K. The peak temperatures are displayed in Fig. \ref{figsize}(b). Note that we should not use the peak heights for finite-size scaling to calculate critical exponents because the discretized temperatures do not allow one to localize the peak height with a high precision. To do that we should use continuous-temperature methods such  as histogram MC techniques which give the exact peak (position and height) at $T_c$ (see for example \cite{Ngo-Diep2007}).

\begin{figure}[ht!]
\centering
\includegraphics[width=7cm,angle=0]{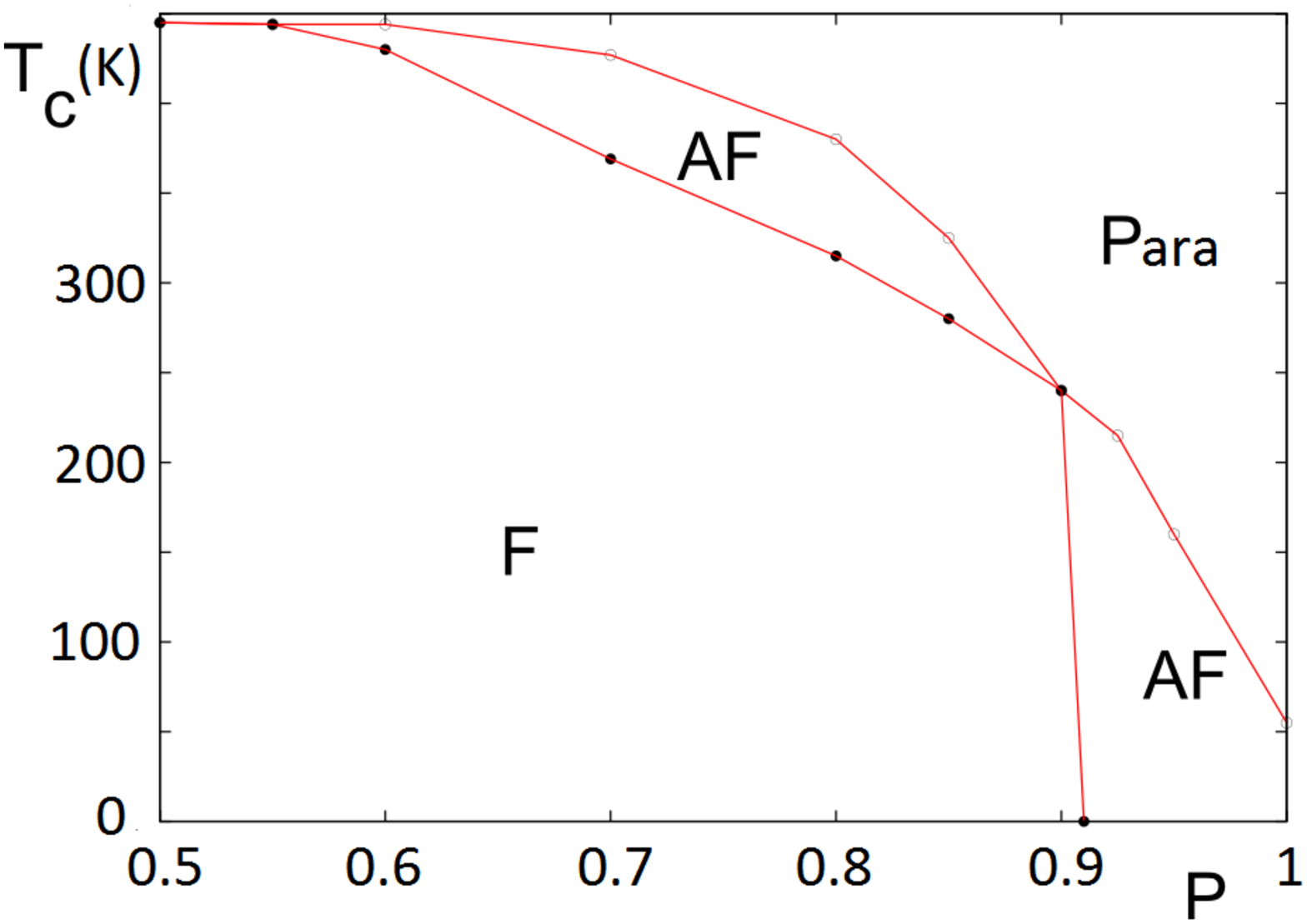}
\caption{(Color online) La$_{0.7}$Sr$_{0.3}$Mn$_{P}^{3+}$Mn$_{1-P}^{4+}$O$_{3}$: Phase diagram in the space ($P,T$) where $P$ is the Mn$^{3+}$ concentration. Black circles denote the ferromagnetic transition temperatures, red void circles indicate the transition from the antiferromagnetic to paramagnetic phase. F, AF and Para denote the ferromagnetic, antiferromagnetic and paramagnetic phases, respectively. One has used $J_1=24.5$ K, $J_2=J_3=-J_4=-0.20$ K. Lines are guides to the eye. See text for comments. \label{figtn}}
\end{figure}

The full diagram is shown in Fig. \ref{figtn}.  This phase diagram is easily understood: near 50\% each of Mn$^{3+}$ is surrounded by Mn$^{4+}$ ions so that antiferromagnetic interactions between Mn$^{3+}$ or between Mn$^{4+}$ are almost absent, there is thus no antiferromagnetic phase. Now,  for larger concentrations near 100\%, the ferromagnetic phase disappears leaving the place for an antiferromagnetic phase because the number of  Mn$^{4+}$ ions is so small, not sufficient to induce a ferromagnetic order: the system is then composed of Mn$^{3+}$ with weak antiferromagnetic interaction $J_2$ between them. This induces an antiferromagnetic phase at very low temperatures.
Let us give the value of the temperature at the ferromagnetic transition for several values of Mn$^{3+}$ concentration $P$: $T_c\simeq 445\pm 5$ K for 50\% and 55\% (large errors are due to large disorder at these concentrations). From 60\% the system undergoes the ferromagnetic transition at  $T_c\simeq 430\pm 3 $ K for 60\%, 369$\pm 1$ K for 70\%, $T_c\simeq 315\pm 1$ K for 80\% and $T_c\simeq 240\pm 1$ K for 90\%.  However, for concentrations from 60\% to 90\%, the antiferromagnetic ordering sets in above $T_c$ up to $T_N\simeq$ 444 K (for 60\%), 427 K (for 70\%) and 380 K (for 80\%). For $P>$90\%, only the antiferromagnetic phase exists.

We emphasize here that the phase diagram in Fig. \ref{figtn} is shown for $J_1=24.5$ K, $J_2=J_3=-J_4=-0.20$ K.  If $J_2$ is smaller, say $J_2=-0.08$ K, the antiferromagnetic phase above the ferromagnetic phase disappears as seen in Fig. \ref{figms002} for $P=70\%$.  We have checked the absence of the antiferromagnetic phase  at other Mn$^{3+}$ concentrations (not shown), except naturally the one in the zone $P \in [0.91,1]$.

At this stage, we would like to recall that the existence of a partially-disordered antiferromagnetic phase above the ferromagnetic phase is
frequently seen  in many frustrated systems \cite{DiepFSS}.  For example, in exactly solved frustrated Ising models such as the
honeycomb lattice \cite{Diep-hc} Kagom\'e lattice \cite{Diep-Kago} among others \cite{DiepSQ}, it has been shown that the transition from the ferromagnetic phase to an antiferromagnetic phase takes place
via a very narrow intermediate paramagnetic phase called ``reentrant phase".  When this reentrance is not possible, the direct transition between the ferromagnetic and the antiferromagnetic phases should be of first-order because one of these two phases is not a symmetry subgroup of the other as in second-order phase transitions. A direct ferromagnetic-antiferromagnetic first-order transition  has been seen in Ref. \onlinecite{Diepbcc}.
The model simulated here has a strong disorder due to the random mixing of Mn$^{3+}$   and Mn$^{4+}$ so that we cannot deal with it analytically, but competing interactions included in the model   (ferromagnetic $J_1$, antiferromagnetic $J_2$ and $J_3$) can cause the ferromagnetic-antiferromagnetic transition. The abrupt change of the total magnetization and the staggered magnetization observed at the transition shown in Figs. \ref{figms} and \ref{figms01} suggest a direct transition with a first-order character.

\subsection{Properties of La$_{0.7}$Sr$_{0.3}$Mn$_{0.7}^{3+}$Mn$_{0.3-x}^{4+}$Ti$_{x}$O$_{3}$}
In this doped material, Ti atoms replace a fraction $x$ of Mn$_{0.3}^{4+}$ with $x$ varying from 0 to $0.3$.  Since Ti is non magnetic, the substitution introduces a dilution.  Let us take into account interactions $J_1$, $J_2$ and $J_3$ between Mn ions as defined above.  In addition, we have added a very small ferromagnetic interaction between Mn$^{3+}$ NNN.

While there is no experimental proof so far on the existence of the antiferromagnetic phase above $T_c$, we choose to study the effect of Ti substitution at $J_2=-0.20$ K hereafter (larger $J_2$ induces an AF phase as seen above).  The agreement with experiments justifies this choice.

\begin{figure}[ht!]
\centering
\includegraphics[width=7cm,angle=0]{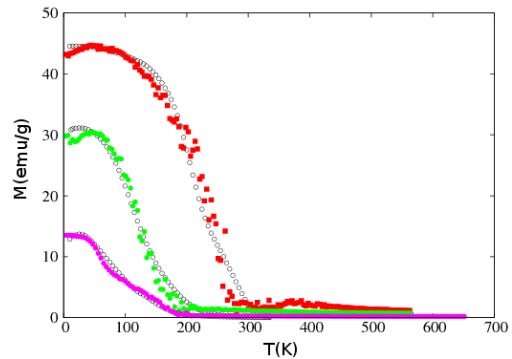}
\caption{(Color online) La$_{0.7}$Sr$_{0.3}$Mn$_{0.7}^{3+}$Mn$_{0.3-x}^{4+}$Ti$_{x}$O$_{3}$: The total magnetization obtained by Monte Carlo simulation  with Ti doping at concentrations (from right to left) $x=0.1$ (red squares), 0.2 (green circles) and 0.25 (violet circles) under applied magnetic field $\mu_0H=0.05$ Tesla are compared with experimental data for $x=0.1$, 0.2 and 0.3 (void black circles) (experimental data taken from Fig. 1 of Ref. \cite{Kallel1}). See text for comments. \label{figtitanm}}
\end{figure}

Using $J_1\simeq 24.5$ K, $J_2=J_3=J_4=\simeq -0.20$ K we have simulated this system with several values of $x$: 0.1, 0.2 and 0.25. The reason why we did not take $x=0.3$ is because at this dilution, there is no more
 Mn$^{4+}$ in the system so that there is no  Mn$^{3+}$- Mn$^{4+}$ interaction, namely no ferromagnetic phase. Experiments have been carried out at this concentration but we believe that there remains a small number of  Mn$^{4+}$ in the material as it has been noticed by Jonker and Van Santen \cite{Jonker}. Not surprisingly, our results for $x=0.25$ agree very well with experimental observations for $x=0.3$.
The full results of the magnetization with applied magnetic field $\mu_0H=0.05$ Tesla are shown in Fig. \ref{figtitanm}   to compare with experiments at the same field \cite{Kallel}.  Let us comment these results:
(i) the overall agreement with experiments is excellent, given the fact that no other parameter adjustments were necessary (we used only exchange parameters while fitting with $T_c$ as described above), (ii) the slight down-turn curvature at very low $T$ is sensitive to $J_2$: smaller $J_2$ will suppress this but then the agreement is less perfect at higher temperatures.

\begin{figure}[ht!]
\centering
\includegraphics[width=7cm,angle=0]{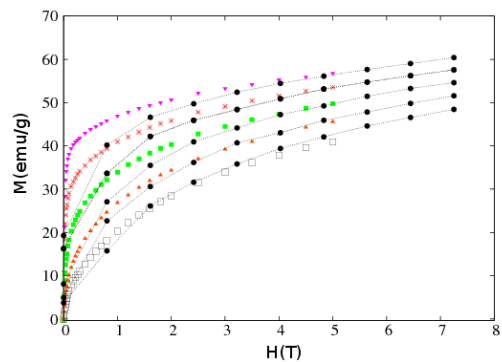}
\caption{(Color online) La$_{0.7}$Sr$_{0.3}$Mn$_{0.7}^{3+}$Mn$_{0.3-x}^{4+}$Ti$_{x}$O$_{3}$: Effect of magnetic field at $x=0.1$. Monte Carlo results (black circles) versus
experimental data at temperatures (from top) 190 K, 200 K, 210 K, 220 K and 230 K (data taken from Fig. 1 of Ref. \cite{Kallel2010}). Lines are guides to the eye. A good agreement is observed except at very low fields. See text for comments. \label{figtitanh}}
\end{figure}

We have also calculated at five temperatures the total magnetization as a function of the applied magnetic field $\mu_0H$.  These results are shown in Fig. \ref{figtitanh} together with experimental data taken from Ref. \onlinecite{Kallel2010}. A good agreement is observed except at very low fields where precision of parameters and experimental details of lattice structure and disorder play certainly an important role.

At this point, we would like to emphasize that doping La$_{0.7}$Sr$_{0.3}$Mn$_{0.7}^{3+}$Mn$_{0.3}^{4+}$O$_{3}$  with other ions such as Ni can alter magnetic structures in a drastic way such as domains with different lattice deformations and creation of ferromagnetic clusters which change the low-$T$ behaviors \cite{Feng}.  We believe that strong Ti-substitution also induces these aspects but probably in a less drastic manner. The good agreement obtained above without introducing such factors may be a proof of the absence of complicated structure deformations.

Before closing this section, let us mention that our present model can also be used to study the case
where a number of La is substituted by magnetic Ce$^{3+}$ ions of spin $S=1/2$:
Ce occupy the centered sites of the bcc lattice. Additional interactions between Ce$^{3+}$ ions
and Mn$^{3+}$ and Mn$^{4+}$ have to be introduced. Experiments have been performed
on La$_{0.56}$Ce$_{0.14}$Sr$_{0.3}$MnO$_{3}$ \cite{Kallel2009,Kallel2010a}. Work is under way to explain by MC simulations magnetic behaviors observed in this system.

\section{Conclusion}\label{sectcon}

Perovskite compounds La$_{0.7}$Sr$_{0.3}$Mn$_{0.7}^{3+}$Mn$_{0.3}^{4+}$O$_{3}$ have very rich magnetic behaviors when substituting Mn ions with other non magnetic atoms. In this paper we have studied this compound without and with Ti substitution by the use of MC simulations.  Fitting only the experimental value of the critical temperature $T_c=369$ K in the case of non substitution, we have estimated various exchange interactions of which the ferromagnetic one between  Mn$^{3+}$ and Mn$^{4+}$ is dominant ($\simeq 24.5$ K).  The other interaction which is though small  but plays an important role is $J_2$ between Mn$^{3+}$ ($\simeq -0.20$ K): it is at the origin of an antiferromagnetic phase observed in a small temperature region above $T_c$. This phase diminishes progressively with decreasing $J_2$ and disappears when $J_2$ becomes smaller (see Fig. \ref{figms002} for $J_2\simeq -0.08$ K, for example). Experiments did not see this phase. There are two possible explanations. The first one is that experiments have overlooked it because experimental techniques used so far were not suitable to detect an antiferromagnetic ordering. The second explanation is that if the value of $J_2$ in experimental samples is smaller than -0.20 K then the antiferromagnetic phase does not exist as shown above.

We have also investigated the effect of Ti substitution on the magnetization as a function of  $T$ at several substitution concentrations. Our results on the magnetization agree remarkably with experiments over the whole range of temperature, showing that the estimated parameters are precise enough to reproduce experimental data at several substitution concentrations (Fig. \ref{figtitanm}).   For the field effect, our results agree with experiments except at very low fields where there may be many other finer aspects such as a possible existence of magnetic domains, impurity clusters, etc. in real crystals which should be taken into account.

\acknowledgments
 SY acknowledges  a financial support from the Ministry of Education of Tunisia. She is grateful to the University of Cergy-Pontoise for hospitality where this work has been accomplished.  The authors are thankful to Dr. A. Ben Lamine for a critical reading of the manuscript.

{}


\begin{thebibliography}{0}%
\makeatletter
\providecommand \@ifxundefined [1]{%
 \@ifx{#1\undefined}
}%
\providecommand \@ifnum [1]{%
 \ifnum #1\expandafter \@firstoftwo
 \else \expandafter \@secondoftwo
 \fi
}%
\providecommand \@ifx [1]{%
 \ifx #1\expandafter \@firstoftwo
 \else \expandafter \@secondoftwo
 \fi
}%
\providecommand \natexlab [1]{#1}%
\providecommand \enquote  [1]{``#1''}%
\providecommand \bibnamefont  [1]{#1}%
\providecommand \bibfnamefont [1]{#1}%
\providecommand \citenamefont [1]{#1}%
\providecommand \href@noop [0]{\@secondoftwo}%
\providecommand \href [0]{\begingroup \@sanitize@url \@href}%
\providecommand \@href[1]{\@@startlink{#1}\@@href}%
\providecommand \@@href[1]{\endgroup#1\@@endlink}%
\providecommand \@sanitize@url [0]{\catcode `\\12\catcode `\$12\catcode
  `\&12\catcode `\#12\catcode `\^12\catcode `\_12\catcode `\%12\relax}%
\providecommand \@@startlink[1]{}%
\providecommand \@@endlink[0]{}%
\providecommand \url  [0]{\begingroup\@sanitize@url \@url }%
\providecommand \@url [1]{\endgroup\@href {#1}{\urlprefix }}%
\providecommand \urlprefix  [0]{URL }%
\providecommand \Eprint [0]{\href }%
\providecommand \doibase [0]{http://dx.doi.org/}%
\providecommand \selectlanguage [0]{\@gobble}%
\providecommand \bibinfo  [0]{\@secondoftwo}%
\providecommand \bibfield  [0]{\@secondoftwo}%
\providecommand \translation [1]{[#1]}%
\providecommand \BibitemOpen [0]{}%
\providecommand \bibitemStop [0]{}%
\providecommand \bibitemNoStop [0]{.\EOS\space}%
\providecommand \EOS [0]{\spacefactor3000\relax}%
\providecommand \BibitemShut  [1]{\csname bibitem#1\endcsname}%
\let\auto@bib@innerbib\@empty
\end{thebibliography}%


\begin{thebibliography}{9}

\bibitem{Zinn} J. Zinn-Justin, {\it Quantum Field Theory and
Critical Phenomena}, Oxford Unversity Press (2002).

\bibitem{DiepSP} H. T. Diep, {\it Statistical Physics: Fundamentals and Application to Condensed Matter}, Word Scientific, New Jersey (2015).

\bibitem{Dagotto2001}  E. Dagotto, T. Hotta, and A. Moreo, Phys. Rep. {\bf 344}, 1 (2001).
\bibitem{Hotta} T. Hotta, A. Feiguin, and E. Dagotto, Phys. Rev. Lett {\bf 86}, 4922 (2001).
\bibitem{Dagotto2000} E. Dagotto, Science 309 (2005) 257, N. Nagaosa and Y. Tokura, Science {\bf 288}, 462 (2000).
\bibitem{Kim} M. S. Kim, J. B. Yang, Q. Cai, X. D. Zhou, W. J. James, W. B. Yelon, P. E. Parris, D. Buddhikot and S. K. Malik, Phys. Rev. B {\bf 71}, 014433 (2005).
\bibitem{Salamon} M. B. Salamon, M. Jaime, The physics of manganites: structure and transport, Rev. Mod. Phys. {\bf 73}, 583 (2001).
\bibitem{Zhu} X. Zhu, U. Yuping, X. Luo, L. Hechang, W. Bosen, S. Wenhai, Y. Zhaorong, D. Jianming, SH. Dongqi, D. Shixue, J. Magn. Magn. Mater. {\bf 322}, 242 (2010).
 \bibitem{Helmolt} R. Von Helmolt, J. Wecker, B. Holzapfel, L. Schultz, K. Samwer, Phys. Rev. Lett. {\bf 71}, 2331 (1993).
\bibitem{Jonker} G. H. Jonker, J. H. Van Santen, Physica {\bf 16}, 337 (1950).
\bibitem{Zener1} C. Zener, Phys. Rev. {\bf 81}, 440 (1951).
\bibitem{Zener2} C. Zener, Phys. Rev. {\bf 82},  403 (1951).
\bibitem{Zener3} C. Zener, Phys. Rev. {\bf 83}, 299 (1951).
\bibitem{Hong}  C. S. Hong, N. H. Hur, Y. N. Choi, Solid State Com. {\bf 131}, 779 (2004).
 \bibitem{Bose}E. Bose, S. Karmakar, B. K. Chaudhuri, S. Pal, Solid State Com. {\bf 145}, 149 (2008).
\bibitem{Furukawa} N. Furukawa, Y. Motome, Appl. Phys. A, {\bf 74}, 1728  (2002).
\bibitem{Urushibara} A. Urushibara, Y. Moritomo, T. Arima, A. Asamitsu, G. Kido, Y. Tokura, Phys. Rev. B {\bf 51}, 14103 (1995).
\bibitem{Hotta1} T. Hotta, E. Dagotto, Phys. Rev. B {\bf 61}R, 11879 (2000).

\bibitem{Restrepo} E. Restrepo-Parra, C. D. Salazar-Enriquez, J. Londono-Navarro, J. F. Jurado, J. Restrepo, J. Magn. Magn. Mater. {\bf 323}, 1477 (2011).
\bibitem{Kallel} N. Kallel, G. Dezanneau, J. Dhahri, M. Oumezzine, H. Vincent, Structure, magnetic and electrical behaviour of La$_{0.7}$Sr$_{0.3}$Mn$_{1–x}$Ti$_x$O$_3$ with $0\le x\le 0.3$, J. Magn. Magn. Mater. {\bf 261}, 56 (2003).
\bibitem{Kallel1} N. Kallel, K. Fr$\ddot{o}$hlich, M. Oumezzine, M. Ghedira, H. Vincent, and S. Pignard, Magnetism and giant magnetoresistance in La$_{0.7}$Sr$_{0.3}$Mn$_{1–x}$M$_x$O$_3$ (M = Cr, Ti) systems, Phys. Stat. Sol. (c) {\bf 1}, 1649–1654 (2004).
\bibitem{Kallel2} N. Kallel, M. Oumezzine, H. Vincent, Neutron-powder-diffraction study of structural and magnetic structure of La$_{0.7}$Sr$_{0.3}$Mn$_{1–x}$Ti$_x$O$_3$ ($x$ = 0, 0.10, 0.20, and 0.30), J. Magn. Magn. Mater. {\bf 320}, 1810 (2008).
\bibitem{Kallel2010} S. Kallel, N. Kallel, O. Pe\u{n}a and M. Oumezzine, Large magnetocaloric effect in Ti-modified La$_{0.70}$Sr$_{0.30}$MnO$_3$ perovskite, Materials Letters {\bf 64}, 1045 (2010).
\bibitem{Goodenough} J. B. Goodenough, A. Loeb, Phys. Rev. {\bf 98}, 391 (1955).
\bibitem{Goodenough1} J. B. Goodenough, {\it Magnetic  Properties of Perovskites}, Londolt-Bornstein Tabellen, Springer-Verlag, Berlin (1962).
\bibitem{Kanamori} J. Kanamori, J. Phys. Chem. Solids {\bf 10}, 87 (1959).
\bibitem{Solovyev} I. Solovyev, N. Hamada, K. Terakura, Phys. Rev. Lett. {\bf 76}, 4825 (1996).
\bibitem{Su} Y. -S. Su, T. A. Kaplan, S. D. Mahanti, J. F. Harrison, Phys. Rev. B {\bf 61}, 1324 (2000).
\bibitem{Birsan} E. B\^{\i}rsan, J. Magn. Magn. Mater. {\bf 320}, 646 (2008).
\bibitem{Anderson} P. W. Anderson, H. Hasegawa, Phys. Rev. {\bf 100}, 675 (1955).
\bibitem{DiepTM} H. T. Diep, {\it Theory of Magnetism: Application to Surface Physics},  World Scientific, New Jersey (2014).
\bibitem{Metropolis} N. Metropolis, A. W. Rosenbluth, M, N, Rosenbluth, and A. H. Teller, J. Chem. Phys. {\bf 21}, 1087 (1953).
\bibitem{Ngo2014} V.-Thanh Ngo, D.-Tien Hoang, H. T. Diep and I. A. Campbell, Effect of Disorder in the Frustrated Ising FCC Antiferromagnet: Phase Diagram and Stretched Exponential Relaxation, Modern Phys. Lett. B {\bf 28}, 1450067 (2014).
\bibitem{EdwardsAnderson}S. F. Edwards and P. Anderson, J. Phys. F {\bf 5}, 965 (1975).

\bibitem{BinderYoung} K. Binder and A. P. Young, Rev. Mod. Phys. {\bf 58}, 801 (1986).
\bibitem{Mezard} M. M\'ezard,  G. Parisi and M. Virasoro, {\it Spin Glass Theory and Beyond},  World Scientific, New Jersey (1987).

\bibitem{Magnin2012} See Y. Magnin and H. T. Diep,  Monte Carlo Study of Magnetic Resistivity in Semiconducting MnTe,
 Phys. Rev. B {\bf 85}, 184413 (2012) and experimental works cited therein.

\bibitem{Ngo-Diep2007}    V. Thanh Ngo and H. T. Diep, Effects of Frustrated Surface in Heisenberg Thin Films,  Phys. Rev. B {\bf 75}, 035412 (2007).

\bibitem{DiepFSS} H. T. Diep (Ed.), {\it Frustrated Spin Systems}, 2nd edition, World Scientific, New Jersey (2013).
\bibitem{Diep-hc} H. T. Diep, M. Debauche and H. Giacomini, Exact solution of an anisotropic centered honeycomb Ising lattice: Reentrance and Partial Disorder, Phys. Rev. B (rapid communication) {\bf 43}, 8759 (1991).

\bibitem{Diep-Kago} M. Debauche, H. T. Diep, H. Giacomini and P. Azaria, Exact Phase Diagram in a generalized Kagom\'e Ising lattice: Reentrance and Disorder Lines, Phys. Rev. B {\bf 44}, 2369 (1991).
\bibitem{DiepSQ} M. Debauche and H. T. Diep,  Successive Reentrances and Phase Transitions in exactly solved dilute centered square Ising lattices,  Phys. Rev. B {\bf 46}, 8214 (1992); H. T. Diep, M. Debauche and H. Giacomini, Reentrance and Disorder Solutions in Exactly Solvable Ising Models,
    J. Mag. Mag. Mater. {\bf 104-107}, 184 (1992).


\bibitem{Diepbcc} P. Azaria,  H. T. Diep and H. Giacomini, First order transition, multicriticality and reentrance in a BCC lattice with Ising spins,  Europhys. Lett. {\bf 9}, 755 (1989).

\bibitem{Feng} J.-W. Feng, C. Ye and L.-P. Hwang, Phys. Rev. B {\bf 61}, 12271 (2000).
\bibitem{Kallel2009} N. Kallel, S. Kallel, O. Pe\u{n}a, M. Oumezzine, J. Solid State Sci. {\bf 11}, 1494 (2009).
\bibitem{Kallel2010a} S. Kallel, N. Kallel, A. Hagaza, O. Pe\u{n}a and M. Oumezzine, J. Alloys and Compounds {\bf 492}, 241 (2010).


\end{thebibliography}
\end{document}